\documentclass[a4paper,12pt,english]{article}

\usepackage[latin2]{inputenc}
\usepackage[most]{tcolorbox}
\usepackage{amsfonts,bm,amssymb,euscript,array,babel,cite}
\usepackage{amsmath,dsfont,bbm}
\usepackage{mathtools}
\usepackage{mathrsfs}
\usepackage{braket}
\usepackage{stmaryrd}
\usepackage{tensor,accents}
\usepackage{hhline}
\usepackage[amsmath]{empheq}
\usepackage{hyperref}
\usepackage{cancel}
\usepackage{multirow}
\usepackage{pdfsync}
\usepackage{tikz}
\usepackage{tikz-cd}
\usepackage{comment}

\numberwithin{equation}{section}

\makeatletter

\newcommand{\dd}{\mathrm{d}}

\newcommand{\sfrac}[2]{{\textstyle\frac{#1}{#2}}}
\def\nn{\nonumber}
\def\mc{\mathcal}
 
\newcommand{\sbullet}{%
  \hbox{\fontfamily{lmr}\fontsize{.8\dimexpr(\f@size pt)}{0}\selectfont\textbullet}}

\makeatother

\def\e{\epsilon} 
   \def\k{\kappa} \def\l{\lambda}
 \def\m{\mu}
\def\n{\nu}  \def\p{\partial}  \def\r{\rho}
\def\s{\sigma}

\def\R{{\mathbb R}}

\begin{document}

\makeatother
\parindent=0cm
\renewcommand{\title}[1]{%
    \vspace{10mm}
    \noindent{\Large{\bf #1}}
    \vspace{8mm}
}
\newcommand{\authors}[1]{%
    \noindent{\large #1}
    \vspace{5mm}
} 
\newcommand{\address}[1]{%
    {\itshape #1\vspace{2mm}}
}

\begin{titlepage}
	
\begin{center}
	
    \title{%
    {\Large Adiabatically-induced Kawaguchi geometry  \\[5pt] and jerk in quantum-classical systems}
    }
	
	\vskip 3mm

 \authors{\large Athanasios Chatzistavrakidis, Larisa Jonke and Ryan Requist
  }
    
  \address{ 
  	 Division of Theoretical Physics, Rudjer Bo\v skovi\'c Institute \\ Bijeni\v cka 54, 10000 Zagreb, Croatia 
  }

  \ 

  \ 

    \begin{abstract}

Adiabatically eliminating the quantum degrees of freedom in a mixed quantum-classical system produces an effective force in the classical equation of motion.  The elimination can be made to any order in the adiabatic parameter, generating a series of higher order forces.  By applying a sequence of near-identity unitary transformations to the quantum state, we derive a hierarchy of increasingly accurate effective actions for the classical variables.  The third order Euler-Lagrange equation is non-Newtonian as the force depends on the jerk, the third order time derivative of position.  
We find that the third order terms induce a special kind of Kawaguchi geometry on the space of classical variables.  This geometry is characterized by an almost symplectic structure and a differential line element that depends on the acceleration in addition to the velocity.  Our results can be used to efficiently capture higher order nonadiabatic effects in molecular dynamics simulations.
    
    \end{abstract}
    
\end{center}

\end{titlepage}

\setcounter{footnote}{0}
\tableofcontents

\newpage 

\section{Introduction} 
\label{sec:introduction}

The fast and slow variables in systems that display dynamics on two vastly different time scales are usually weakly coupled.  But weak interactions can add up to a significant effect if they act over a long time, as illustrated by the synchronization of Huygens' clocks or the slow rotation of the swing plane in Foucault's pendulum.  Although the dynamics of such classical fast-slow systems could be fully described by integrating Newton's second law, in practice this is challenging because it requires integrating over many periods of the fast oscillation to reach long times.  Perhaps more importantly, such a calculation would not explain the slow dynamics in simple terms.  Why do the clocks synchronize?  Why does the pendulum rotate?

Such questions can be answered by deriving a reduced description of the slow variables, usually as a closed set of effective equations of motion.  The elimination of the fast variables can be achieved, in some cases, by assuming that they adapt instantaneously to changes in the slow variables \cite{vankampen1985}. This amounts to treating the fast variables as functions of the slow variables and using those functions to eliminate the fast variables.  The ratio of the fast and slow time scales, when treated as a tunable adiabatic parameter, can be used to specify the accuracy of asymptotic approximations in the adiabatic limit.

Effective equations of motion often have geometrical significance.  For example, the effective equation of motion for a vector defining the slowly-changing swing plane of the Foucault pendulum is equivalent to the parallel transport law on a sphere.  Geometry is known to emerge from low order adiabatic approximations, but the geometry of higher order approximations is almost completely unexplored. 

Interesting geometric structures arise in quantum systems.  Mead and Truhlar found that the effective nuclear Schr\"odinger equation of a molecule contains an induced gauge potential that depends on the state of the electrons, assumed to remain in the same Born-Oppenheimer eigenstate for all values of the nuclear positions \cite{mead1979}.  The Mead-Truhlar gauge potential causes interference phenomena as well as qualitative changes in the energy eigenvalues of electronic-vibrational states \cite{longuet-higgins1958,ham1987,mead1992}. 
Additional geometric structures in the effective nuclear Schr\"odinger equation, such as an induced scalar potential depending on a Riemannian metric \cite{provost1980} on the manifold of nuclear coordinates, have been studied\cite{jackiw1988,berry1989,berry1990,berry1993}. In time-dependent quantum systems driven by slowly changing external parameters, the lowest order adiabatic wave function contains a geometric phase, the Berry phase, in addition to the dynamical phase \cite{berry1984}.

The present study focuses on mixed quantum-classical dynamics.  A paradigmatic example is molecular dynamics, where electrons are described quantum mechanically and nuclei are treated as classical particles moving along a trajectory $x=x(t)$.  Although our results apply to general quantum-classical dynamics, for concreteness we consider Ehrenfest molecular dynamics  \cite{bornemann1996,prezhdo1997,tully1998,todorov2001,alonso2011}.  The electronic state $|\psi(t)\rangle$ evolves according to a time-dependent Schr\"odinger equation with a Hamiltonian $H(x)$ that depends parametrically on the nuclear coordinates $\{x^{\m}\}$.  The nuclei obey Newton's law with the force $-\langle \psi | \partial_{\m} H |\psi\rangle$.

We introduce a method for deriving an effective equation of motion for the nuclei that is accurate to any desired order in a small parameter $\e$ that controls the rate of change of $x$, i.e.~$\dd x^{\m}/\dd t$ is assumed to be proportional to $\e$.  As the nuclei move along a trajectory, the electrons experience a slowly-varying time-dependent Hamiltonian.  The adiabatic theorem implies that if the electronic state is initially in the $n$th eigenstate $\ket{n(x_0)}$ of the Hamiltonian $H(x_0)$, then at a later time it is approximately $\ket{\psi(t)}=(\textrm{phase factor})\ket{n(x)}$, where $\ket{n(x)}$ is the instantaneous eigenstate at $x=x(t)$ \cite{born1928, kato1950}.  In this zeroth order approximation, the nuclei feel the adiabatic force $-\partial_{\m} E_n$ of Born-Oppenheimer molecular dynamics.  The electronic variables have been eliminated; their only role is to shape the adiabatic potential energy surface $E_n(x)$.

The zeroth order approximation is purely adiabatic, as it contains no contribution from states other than $\ket{n(x)}$.  Higher order corrections must account for the admixture of other states due to nonadiabatic coupling.  This admixture renormalizes the effective force, which slightly alters the nuclear trajectory, in turn further modifying the electronic state, and so on {\it ad infinitum}.  We aim to unspool this feedback mechanism order-by-order in $\e$. 

To find higher order corrections, we derive a hierarchy of increasingly accurate effective actions. The corresponding equations of motion produce successively higher order corrections to the adiabatic reaction force acting on the nuclei.  Other approaches to calculating arbitrarily high order adiabatic reaction forces have been used in fast-slow systems in which both components are classical \cite{berry2010,littlejohn1993,weigert1993} or systems in which one component is classical and the other is treated semiclassically \cite{littlejohn1993,weigert1993}.

The first order force is an effective Lorentz force with the Berry curvature acting as a generalized magnetic field.  It is used in adiabatic molecular dynamics simulations in an external magnetic field  \cite{ceresoli2007,culpitt2021}.   

The second order force, being of second order in time derivatives, can be absorbed into the mass-times-acceleration term of Newton's law by defining a covariant acceleration and renormalizing the nuclear masses. The resulting second order equation of motion has the same Newtonian form as the zeroth and first order equations, but the bare nuclear masses get replaced by an effective mass tensor $M^{\star}_{\m\n}$. The effective mass tensor provides a direct link between geometry and dynamics: for a free particle, the second order equation of motion is simply the geodesic equation corresponding to $M^{\star}_{\m\n}$ \cite{goldhaber2005}. 

Our main result is the evaluation of the third order force, which brings about a fundamental change. Because it contains the third order time derivative $\dd^3 x^{\m}/\dd t^3$, known as {\it jerk}, it changes the effective equation of motion from a second order to a third order differential equation. This raises an important conceptual question: how should the jerk be defined as a covariant quantity? We address this question by identifying induced third order geometric structures and showing that they supply a connection that can be used to define the jerk as the covariant derivative of the acceleration.

We isolate the third order terms in the equation of motion and study their geometry for its own sake. All of the third order terms can be generated by an effective Lagrangian $L_3$.  Its Euler-Lagrange equation is equivalent to the equation defining a generalized geodesic in a topological space called Kawaguchi space. In a Kawaguchi space, the infinitesimal line element depends not only on the velocity but also on higher order derivatives, such as the acceleration \cite{Kawaguchi}. Unless a Kawaguchi space is endowed with additional geometric structure, it is not possible to define the length of a tangent vector and the angle between two tangent vectors.  The Kawaguchi space corresponding to $L_3$ is of a special class studied in Refs.~\cite{Losik,Panzhenskii}.  It is equipped with an almost symplectic two-form $\omega$ (a nondegenerate two-form that is not necessarily closed) and an additional geometric structure that gives rise to a connection. For a free particle moving along a Kawaguchi geodesic, we show that conservation of energy is equivalent to the conservation of the (almost) symplectic area $\omega_{\m\n}v^{\m} a^{\n}$ between the velocity and acceleration vectors. This is the counterpart of the conservation of the kinetic energy $\frac{1}{2} M_{\m\n}v^{\m} v^{\n}$ of a free particle in a Riemannian space.

\section{Mixed quantum-classical dynamics} 
\label{sec:CQ}

We are interested in systems comprising some degrees of freedom that can be treated classically and some that are treated quantum mechanically. A typical example is a system of electrons and nuclei. The electrons are described by a time-dependent state $\ket{\psi}=\ket{\psi(t)}$, whereas the nuclei are treated classically and their collective coordinates are denoted $x^{\mu}=x^{\mu}(t)$, with $\mu=1,\dots,N$. The electronic Hamiltonian  $H=H(x)$ depends on the $N$ independent parameters $x^{\mu}$, which can be thought of as coordinates of a manifold $\mc{M}$ of dimension $N$. Such systems can be described exactly as Lagrangian mechanical systems with an action functional $S$ that depends on the electronic state and the $N$ real components $x^{\mu}$ of a map $x: \Sigma_{t}\to \mc{M}$, where $\Sigma_{t}$ is the worldline. Its general form is 
\begin{equation}\label{eq:bare action}
    S[x,\ket{\psi}]=\int\dd t \left[\frac 12 M_{\mu\nu}(x)\, \dot{x}^{\mu}\dot{x}^{\nu}+\bra{\psi}i \frac{\dd}{\dd t} -H(x)\ket{\psi}\right]\,,
\end{equation}
where $M_{\mu\nu}=M_{\mu\nu}(x)$ is the position-dependent bare mass tensor for the nuclei, a positive-definite, symmetric bilinear form (a Riemannian metric on the parameter manifold $\mc{M}$), and $\dot{x}^{\m}=\dd x^{\m}/\dd t$.  We use atomic units so that $\hbar=1$.

The Euler-Lagrange equations of the action functional in Eq.~\eqref{eq:bare action} are Newton's second law and the time-dependent Schr\"odinger equation, and they read
\begin{subequations} \label{eq:Mean field dynamics}
\begin{align}
\label{eq:Newton law}    M_{\mu\nu}\,a^{\nu}&=-\bra{\psi}\partial_{\mu}H\ket{\psi}\,, \\[4pt] 
  \label{eq:Schrodinger time dependent} 
  i\frac{\dd}{\dd t}\ket{\psi}&=H(x)\ket{\psi}\,.
\end{align}
\end{subequations}
On the left-hand side of Eq.~\eqref{eq:Newton law}, $a^{\nu}$ is the acceleration vector, which on a Riemannian manifold is defined as 
\begin{equation}
\label{eq:acceleration}
a^{\mu}=\ddot{x}^{\mu}+\mathring\Gamma^{\mu}_{\nu\rho}[M]\dot{x}^{\nu}\dot{x}^{\rho}\,,
\end{equation}
where $\mathring\Gamma^{\mu}_{\nu\rho}[M]$ are the coefficients of the Levi-Civita connection $\mathring\nabla[M]$ with respect to the metric $M_{\mu\nu}$, namely
\begin{equation}
    \mathring\Gamma^{\mu}_{\nu\rho}[M]=\frac 12 \,M^{\mu\sigma} \left(\partial_{\nu}M_{\sigma\rho}+\partial_{\rho}M_{\sigma\nu}-\partial_{\sigma}M_{\nu\rho}\right)\,,
\end{equation}
using the inverse of $M_{\mu\nu}$ with components $M^{\mu\nu}$.
Recall that the Levi-Civita connection is the unique symmetric connection which is compatible with a given Riemannian metric. Since in the following we are going to encounter such connections for effective metrics other than the bare mass tensor, we use the following notation. We denote the Levi-Civita connection for a metric $g$ as $\mathring\nabla[g]$ and its coefficients in a given coordinate basis as $\mathring\Gamma^{\mu}_{\nu\rho}[g]=\mathring\Gamma^{\mu}_{\rho\nu}[g]$. More generally, we will denote any symmetric, i.e.~torsion-free, connection as $\mathring\nabla$ even when it is not associated with a metric.

The term on the right-hand side of Eq.~\eqref{eq:Newton law} is the force; when absent, the equation of motion is the geodesic equation for free particles in curved space. 
Equations~\eqref{eq:Mean field dynamics} are the equations of Ehrenfest molecular dynamics describing the electronic and nuclear evolution. As described in the Introduction, the objective is to reduce these two equations to a single, effective classical equation for the nuclear position, thereby eliminating the electronic degrees of freedom.  As well as providing insights into the dynamics of the nuclei, this reduction has practical implications for molecular dynamics simulations, namely it lets us incorporate some nonadiabatic effects while avoiding the computationally demanding task of solving the electronic Schr\"odinger equation, which typically requires a much smaller time step than the nuclear equation.  There are several alternative molecular dynamics methods, including a method that uses a Hilbert space formulation of classical mechanics and is able to reproduce some results of fully quantum simulations \cite{bauer2024}.

\section{Adiabatic perturbation theory}
\label{sec:APT}

Adiabatic perturbation theory is a method capable of approximating quantum systems with slowly-varying time-dependent Hamiltonians.  Its different versions can be used to derive approximate solutions to the Schr\"odinger equation or effective Hamiltonians.
The electronic Hamiltonian $H(x)$ in Eq.~\eqref{eq:Schrodinger time dependent} is slowly varying if the nuclear coordinates $x^{\m}$ change slowly in time. 
To control the time scale of the nuclear motion, we will introduce an adiabatic parameter $\e$ such that the nuclear velocities are proportional to $\e$.  Since this lets us control the rate of change of $H(x)$,
adiabatic perturbation theory can be used to derive approximate solutions 
that have a specific order of accuracy with respect to $\e$ in the adiabatic limit $\e\rightarrow 0$.
We will say that an approximate solution $\ket{\psi^{\scriptscriptstyle{{\rm approx},p}}}$ is a $p$th order adiabatic approximation to the actual solution $\ket{\psi}$ if 
\begin{equation}
    \lim_{\epsilon\to 0}\, \frac 1{\epsilon^{p}}\, ||\ket{\psi^{\scriptscriptstyle{{\rm approx},p}}}-\ket{\psi}||=0\,
\end{equation}
for all times in a finite interval $[t_0,t_1]$.  Since the error vanishes faster than $\e^p$, we say that the solutions obtained from adiabatic perturbation theory have controlled accuracy.

The reason that nuclei, under the assumption that they are treated classically, usually evolve on a slower time scale than electrons is a consequence of their large masses as well as the choice of initial conditions.  The initial state of the electrons is typically chosen to be an adiabatic eigenstate, e.g.~$\ket{\psi(t_0)}=\ket{n(x_0)}$, so that the initial conditions for the nuclear coordinates $x_0^{\m}=x^{\m}(t_0)$ and velocities $v_0^{\m} = \dot{x}^{\m}(t_0)$ correspond to a certain amount of excess energy with respect to the minimum of the adiabatic potential $E_n(x)$ [defined in Eq.~\eqref{eq:adiabatic states}].
The magnitude of this excess energy is usually independent of the nuclear mass scale $M$ or only weakly dependent on it.
We further assume that the electronic eigenvalue $E_n(x)$ is isolated from all other eigenvalues in the spectrum of $H(x)$ by a finite energy gap. 
Given these assumptions, the nuclear velocities are of the order $M^{-1/2}$.  
This motivates us to introduce a small dimensionless parameter $\e$ and scale the nuclear mass tensor according to $M_{\m\n} \rightarrow \e^{-2} M_{\m\n}$.
After this scaling the nuclear velocities become proportional to $\e$, and
the nuclear equation of motion, Eq.~\eqref{eq:Newton law}, becomes
\begin{align}
\label{eq:Newton law:scaled mass}    
\e^{-2} M_{\mu\nu}\,\bigg[\frac{\dd^2 x^{\nu}}{\dd t^2}+\mathring{\Gamma}^{\n}_{\r\s}[M] \frac{\dd x^{\r}}{\dd t} \frac{\dd x^{\s}}{\dd t}\bigg]&=-\bra{\psi}\partial_{\mu}H\ket{\psi}\,.
\end{align}
Changing the time coordinate to $t'=\e t$ removes the explicit appearance of the $\e^{-2}$ factor.  The time coordinate $t'$, called the slow time, is the natural time scale adapted to the nuclear motion.  Under this change of time coordinate, Eqs.~\eqref{eq:Newton law:scaled mass} and \eqref{eq:Schrodinger time dependent} become
\begin{subequations} \label{eq:Mean field dynamics:scaled time}
\begin{align}
\label{eq:Newton law:scaled time}    
M_{\mu\nu}\,\bigg[\frac{\dd^2 x^{\nu}}{\dd t'^2}+\mathring{\Gamma}^{\n}_{\r\s}[M] \frac{\dd x^{\r}}{\dd t'} \frac{\dd x^{\s}}{\dd t'}\bigg]&=-\bra{\psi}\partial_{\mu}H\ket{\psi}\,.\\[4pt]
\label{eq:Schrodinger time dependent:scaled time}    
    i \,\epsilon\, \frac{\dd}{\dd t'}\ket{\psi}=H(x)\ket{\psi}\,.
\end{align}
\end{subequations}
Equation \eqref{eq:Schrodinger time dependent:scaled time} is the standard form of adiabatic perturbation theory.  
It can be identified as a singularly-perturbed differential equation due to the appearance of $\e$ multiplying the time derivative.  

From now on, we think of Eqs.~\eqref{eq:Newton law:scaled time} and \eqref{eq:Schrodinger time dependent:scaled time} as the starting point for an asymptotic analysis with the adiabatic parameter $\e$ serving as a bookkeeping parameter that identifies different orders of perturbation theory.  Hence, we drop the prime on $t'$.
The adiabatic parameter can be set to unity at the end of the calculation.  

We will use $\e$ to classify the order of accuracy of several quantities, including the approximate wave function, the effective force, the effective Hamiltonian, and the effective action.  The orders of different quantities are usually in correspondence, but there are exceptions.  For example, the $p$th order effective electronic Hamiltonian gives the $(p+1)$th order effective action.

The zeroth order adiabatic approximation to Eq.~\eqref{eq:Schrodinger time dependent:scaled time},
\begin{equation}
\label{eq:psi order 0}
\ket{\psi^{\scriptscriptstyle{{\rm approx},0}}}=e^{i\epsilon^{-1}\phi(t)} e^{i\gamma(t)} \ket{n},
\end{equation}
was derived by Berry \cite{berry1984}.
The state $\ket{n}=\ket{n(x)}$ is the normalized $n$th eigenstate of the Hamiltonian $H=H(x)$, i.e.
\begin{equation}
\label{eq:adiabatic states}
    H(x)\ket{n}=E_{n}(x)\ket{n}\,, \quad \braket{n|n}=1\,.
\end{equation} 
The eigenstates $\ket{n}$ will be called adiabatic eigenstates. The zeroth order adiabatic solution contains the dynamical and geometric phases defined, respectively, as
\begin{equation}
    \phi(t)=-\int_{t_0}^{t}\dd s \,E_{n}(x(s)) \qquad \text{and} \qquad \gamma(t)=\int_{t_0}^{t}\dd s\, A_{\mu}\frac {\dd x^{\mu}}{\dd s}\,.
\end{equation}
In the present context, the geometric phase depends on the components
\begin{equation}
    \label{eq:Berry connection} 
    A_{\mu}= i\braket{n|\partial_{\mu}n}\,
\end{equation}
of the Berry-Mead-Truhlar connection, which is a connection on a principal circle bundle over the parameter manifold $\mc{M}$. $A_{\m}$ could be written as $A_{\m,nn}$ to indicate that it is specific to the state $\ket{n}$, yet we will sometimes suppress the subscript $n$ on $A_{\m,nn}$ and on related quantities for brevity.  The curvature of the Berry-Mead-Truhlar connection is defined as 
\begin{equation}
    B_{\mu\nu}=\partial_{\mu}A_{\nu}-\partial_{\nu}A_{\mu}\,.
\end{equation}
The choice of a connection splits the tangent space of the principal bundle at a point into vertical (tangent to the fibre) and horizontal subspaces, such that the change of the state is separated into 
\begin{equation}
    \partial_{\mu}\ket{n}=\partial_{\mu}\ket{n}|_{\mathrm{hor}}+\partial_{\mu}\ket{n}|_{\mathrm{vert}}\,,
\end{equation}
where the horizontal and vertical components are, respectively,
\begin{equation}
    \partial_{\mu}\ket{n}|_{\mathrm{hor}}= (\partial_{\mu}+iA_{\mu})\ket{n}\,, \qquad \partial_{\mu}\ket{n}|_{\mathrm{vert}}=\ket{n}\braket{n|\partial_{\mu}n}\,.
\end{equation}
The horizontal part is the gauge covariant derivative, which means it transforms covariantly under Abelian gauge transformations (changes of phase): 
\begin{align}
 &   D_{\mu}=\partial_{\mu}+iA_{\mu}\,,\\[4pt] 
&    \ket{n(x)}\mapsto e^{i\theta(x)}\ket{n(x)} \,\Rightarrow D_{\mu}\ket{n(x)}\mapsto e^{i\theta(x)}D_{\mu}\ket{n(x)}\,.
\end{align}
This horizontal part, which satisfies the orthogonality condition
\begin{equation}\label{eq: orthogonality}
    \braket{n|D_{\mu}n}=0\,,
\end{equation} 
corresponds to the physical change in the state during motion in the parameter space, whereas the vertical part is a pure gauge and therefore it corresponds to a redundancy in our description of the state of the system. Since physical states live in the ray space, the projective Hilbert space{\footnote{We mainly refer to the case of finite-dimensional Hilbert space $\mc{H}\simeq \mathbb{C}^{d}$ of (complex) dimension $d\ge N$, in which case the projective Hilbert space $\mathbb{P}\mc{H}$ has (complex) dimension $d-1$. What follows also applies, with some caveats, to the infinite-dimensional case. We will not delve into the subtleties of this here.}} $\mathbb{P}\mc{H}$, and phases are physically irrelevant, removing phases directly leads to a metric, which measures the distance between nearby physical states. This is the quantum geometric tensor 
\begin{equation}
    h_{\mu\nu}=\braket{D_{\mu}n|D_{\nu}n}=g_{\mu\nu}-\frac i2 B_{\mu\nu}\,,
\end{equation}
whose real part is a Riemannian metric $g_{\mu\nu}$ \cite{provost1980} and whose imaginary part is (proportional to) the Berry curvature tensor.{\footnote{More precisely, the real part does not always have to be a Riemannian metric, in particular it may exhibit degeneracy. We assume nondegeneracy at this point.}} To be precise, the projective Hilbert space is canonically equipped with a K\"ahler metric, the Fubini-Study metric $h^{\scriptscriptstyle{FS}}$, whose line element is given as 
\begin{equation}
     \dd s_{\scriptscriptstyle{FS}}^2= \braket{\dd\psi|\dd\psi}-|\braket{\psi|\dd\psi}|^2\,.
\end{equation}
This metric is intrinsic and does not depend on any choices (of coordinates or of connection.) On the other hand, the quantum geometric tensor is induced and it is obtained as the pullback of the Fubini-Study metric,
\begin{equation}
     h_{\mu\nu}=\psi^{\ast}(h^{\scriptscriptstyle{FS}})_{\mu\nu}\,,
\end{equation}
where the pullback is taken with respect to the map 
\begin{align}
\label{eq:map}
     \psi: \mc{M}&\to \mathbb{P}\mc H \\[4pt] 
     x &\mapsto [\ket{\psi(x)}]
\end{align}
from the parameter manifold to the projective Hilbert space and $[-]$ denotes the equivalence class of rays. In the maximal case, namely when the dimension of the (finite-dimensional) parameter manifold equals the dimension of the projective Hilbert space, there is little difference between the two metrics and they can be identified after a choice of coordinates. On the other hand, when the number of parameters on which the Hamiltonian depends is less than the dimension of the projective Hilbert space, the distinction is important, since the two metrics may have different properties, for example the quantum geometric tensor might not be a K\"ahler metric.  

The zeroth order solution in Eq.~\eqref{eq:psi order 0} is purely adiabatic, since it has no contribution from any state other than $\ket{n}$.  
To higher orders in $\e$, we expect the approximate solution to acquire a series of nonadiabatic corrections, $\e\ket{n_1}$, $\e^2\ket{n_2}$, etc., accounting for the admixture of other states. 
Adiabatic perturbation theory was used to derive a $p$th order adiabatic solution in the general form \cite{Requist:2022vox}
\begin{equation}\label{eq:psi p-order}
    \ket{\psi^{\scriptscriptstyle{{\rm approx},p}}}=e^{i\epsilon^{-1}\phi}e^{i\gamma}e^{i\epsilon\alpha_1}\dots e^{i\epsilon^{p}\alpha_{p}}\left(\ket{n}+\epsilon\,\ket{n_1}+\dots + \epsilon^{p}\,\ket{n_p}\right)\,.
\end{equation}
The eigenstate corrections $\ket{n_k}$ are time-local functions of the path $x(t)$ in parameter space, while the $\alpha_k=\alpha_k(t)$ are higher order geometric phases.  In this notation, the Berry phase $\gamma$ corresponds to $\alpha_0$.  The geometrical and dynamical significance of the $\alpha_k$'s will be discussed in the next section.

The approximate solution $\ket{\psi^{\scriptscriptstyle{{\rm approx},p}}}$
can be used to evaluate the force in Eq.~\eqref{eq:Newton law:scaled time} order-by-order in $\e$.  We have carried out this calculation to third order in $\e$.  Interestingly, the resulting equation of motion for the coordinate $x$ appears to be non-Lagrangian, i.e.~we have not been able to obtain it as the Euler-Lagrange equation of any Lagrangian.  Although this is not in itself problematic, e.g.~the equation is solvable and conserves energy, there are advantages to working with equations that can be derived from a Lagrangian.

For this reason, we aim to derive a hierarchy of increasingly accurate effective actions for the coordinate $x$.  The Euler-Lagrange equations will provide a hierarchy of effective equations of motion.
Since we will show in the next section that the $p$th order effective action can be obtained directly from a $(p-1)$th order effective electronic Hamiltonian, we devote the remainder of this section to deriving a hierarchy of effective electronic Hamiltonians. 

We begin by defining a unitary transformation $U_0$ that diagonalizes the electronic Hamiltonian $H(x)$, i.e.
\begin{align}
U_0^{\dag} H U_0 = E = \textrm{diag}(E_1,E_2,\ldots)\,{.}
\end{align}
The unitary operator $U_0$ transforms from any basis $\{|a\rangle\}$ in which the electronic Hamiltonian was originally stated to the basis of adiabatic electronic states denoted by $|n\rangle$, with the matrix element $U_{0,an}=\langle a|n\rangle$.

Defining $|\psi^{(0)}\rangle := U_0^{\dag} |\psi\rangle$, the Schr\"odinger equation in the adiabatic representation is
\begin{align}
i \e \partial_t |\psi^{(0)}\rangle = H^{(0)} |\psi^{(0)}\rangle {}
\label{eq:schroedinger:0}
\end{align}
with the Hamiltonian
\begin{align}
H^{(0)} &= E - i \e U_0^{\dag} \dot{U}_0\,.
\end{align}
We will use the notation 
\begin{align}
F = U_0^{\dag} \dot{U}_0\,,
\end{align}
with the matrix element $F_{ln}=\langle l|\partial_t n\rangle$.
Equation \eqref{eq:schroedinger:0} is equivalent to the original Schr\"odinger equation in Eq.~\eqref{eq:Schrodinger time dependent:scaled time}.  It is merely expressed in a specific (time dependent) representation.  Since we will be working with a hierarchy of adiabatic representations, we use a superscript, e.g.~$(0)$, to indicate the order in the hierarchy.  The usual adiabatic representation, in which states and operators are expressed in the basis of the adiabatic states, is what we now call the zeroth order adiabatic representation.
We define the effective Hamiltonian at this order to be the diagonal part of $H^{(0)}$, thereby neglecting the nonadiabatic coupling, and write
\begin{align}
\label{eq:Heff0}
H^{\rm eff(0)} 
&= E - \e \sum_n A_{\m,nn} \dot{x}^{\m} |n\rangle\langle n|\,{.}
\end{align}
The zeroth order effective Hamiltonian depends on $x$ and $\dot{x}$.
The solution of the effective Schr\"odinger equation with this Hamiltonian and the initial condition $\ket{\psi(t_0)}=\ket{n(x(t_0))}$ reproduces Berry's result in Eq.~\eqref{eq:psi order 0}.

Now we seek higher order corrections.  Consider the near-identity unitary transformation
\begin{align}
\label{eq:U1}
U_1 = e^{i\e G_1} {}
\end{align}
with the Hermitian generator $G_1$.  The unitary operator $U_1$ will be constructed to transform from the zeroth order adiabatic representation to a first order adiabatic representation in which the off-diagonal elements of the new Hamiltonian are an order of magnitude smaller with respect to $\e$.  To see how to construct $G_1$, we expand the new Hamiltonian 
\begin{align}
H^{(1)} = U_1^{\dag} E U_1 - i \e U_1^{\dag} F U_1 - i \e U_1^{\dag} \dot{U}_1 
\end{align}
in powers of $\e$ with the help of a Baker-Campbell-Hausdorff-type formula, obtaining
\begin{align}
H^{(1)} &= E - i \e [G_1, E] + \frac{1}{2!} (-i\e)^2 [G_1, [G_1, E]] + \frac{1}{3!} (-i\e)^3 [G_1,[G_1, [G_1, E]]] \nonumber \\
& -i \e F + (- i \e)^2 [G_1, F] + \frac{1}{2!} (-i\e)^3 [G_1, [G_1, F]] + \e^2 \dot{G}_1 - \frac{i}{2} \e^3 [G_1,\dot{G}_1] + \mathcal{O}(\e^4)\,.
\end{align}
The order $\e$ off-diagonal part can be made to vanish by requiring $G_1$ to satisfy
\begin{align}
G_{1,ln}=\frac{F_{ln}}{E_l-E_n}\quad \mathrm{for}\; l\neq n~, 
\label{eq:condition:1}
\end{align}
and we choose the diagonal elements of $G_1$ to vanish, which is a gauge choice.

Using the condition in Eq.~(\ref{eq:condition:1}) to eliminate $[G_1,E]$ in $H^{(1)}$, we obtain  
\begin{align}
H^{(1)} 
&= E -i\e F_{\rm diag} - \frac{1}{2} \e^2 [G_1,F_{\rm off\, diag}] -\e^2 [G_1,F_{\rm diag}] + \e^2 \dot{G}_{1} \nn \\
&\quad+ \frac{i}{3} \e^3 [G_1,[G_1,F_{\rm off\,diag}]] + \frac{i}{2} \e^3 [G_1,[G_1,F_{\rm diag}]]  -\frac{i}{2} \e^3 [G_1,\dot{G}_1]\,{,}
\end{align}
where $F_{\rm diag} = \sum_n F_{nn} |n\rangle \langle n|$.

The first order effective Hamiltonian is defined to be the diagonal part of $H^{(1)}$, truncated to order $\e^2$. To obtain it, we evaluate
\begin{align}
-\frac{1}{2} [G_1,F_{\rm off\, diag}]_{nn} &= -\frac{1}{2} M_{2\m\n} \dot{x}^{\m} \dot{x}^{\n}\,{,}
\label{eq:second-order:terms}
\end{align}
where 
\begin{align}
M_{2\m\n} &= -2 \mathrm{Re} \langle D_{\m} n| (E_n-H)^{-1} |D_{\n} n\rangle {}
\end{align}
is a tensor induced on parameter space that will be shown to renormalize the bare mass tensor $M_{\m\n}$.  It is a nonadiabatic quantity in the sense that, unlike the quantum geometric tensor, it is not expressible solely in terms of the intrinsic geometry associated with $|n\rangle$.  Instead, it is sensitive to e.g.~the energy gaps $E_n-E_m$ to other states.  The induced mass term has appeared in a variety of contexts and has been derived by several different methods \cite{littlejohn1993,weigert1993,panati2003,goldhaber2005,requist2010,scherrer2017,matyus2019,Requist:2022vox,littlejohn2024,ren2026}.

Since the diagonal parts of the $-\e^2 [G_1,F_{\rm diag}]$ and $\e^2 \dot{G}_{1}$ terms vanish, the general diagonal element of the first order effective Hamiltonian is 
\begin{align}
\label{eq:Heff1:nn}
H_{nn}^{\rm eff(1)} = E_n -\e A_{\m,nn}\dot{x}^{\m}-\frac{\e^2}{2} M_{2\m\n} \dot{x}^{\m} \dot{x}^{\n}\,{.}
\end{align}
This element determines the time-dependent phase of the solution
\begin{align}
\label{eq:psi:eff:1}
\ket{\psi^{\rm eff(1)}} = e^{-i\e^{-1}\int_{t_0}^t H_{nn}^{\rm eff(1)}(s) ds} |n^{(1)}\rangle
\end{align}
of the effective Schr\"odinger equation in the first order adiabatic representation; $|n^{(1)}\rangle$ denotes the $n$th adiabatic eigenstate in this representation.  Due to the choice $G_{1,nn}=0$, this is in fact the same phase that appears in the approximation wave function in Eq.~\eqref{eq:psi p-order}. Indeed, the coefficient of the order $\e^2$ term in Eq.~\eqref{eq:Heff1:nn} is equal to $-\dot{\alpha}_1$, i.e.~the time derivative of the nonintegrable phase $\alpha_1$, as found by a different method in Ref.~\cite{Requist:2022vox}. In the following section, we will see that the time-dependent phase serves as an effective action with which to derive an effective equation of motion for $x$.

Continuing to second order, we consider the unitary transformation
\begin{align}
\label{eq:U2}
U_2 = e^{i\e^2 G_2}\,{.}
\end{align}
The Hamiltonian in the second order adiabatic representation is
\begin{align}
H^{(2)} = U_2^{\dag} H^{(1)} U_2 -i\e U_2^{\dag} \dot{U}_2\,.
\end{align}
The generator $G_2$ is chosen such that the off-diagonal part of $H^{(2)}$ vanishes to order $\e^2$: 
\begin{align}
G_{2,ln}  =- \frac{i}{E_l-E_n}\big(\sfrac{1}{2} [G_1,F_{\rm off\, diag}]_{ln} +[G_1,F_{\rm diag}]_{ln} -\dot{G}_{1,ln}\big) \quad \mathrm{for} \quad l \neq n\,.
\end{align}
With this choice for $G_2$, the order $\e^0$, $\e^1$, and $\e^2$ terms of $H^{(2)}$ are the same as in $H^{(1)}$. 

The third order contributions are
\begin{align}
H^{(2)}|_{\text{third order}}=\frac{i}{3} \e^3 [G_1,[G_1,F_{\rm off\,diag}]] +  \frac{i}{2}\e^3 [G_1,[G_1,F_{\rm diag}]] - \frac{i}{2} \e^3 [G_1,\dot{G}_1] +\e^3 \dot{G}_2\,.
\end{align}
The second order effective Hamiltonian is defined to be the diagonal part of $H^{(2)}$ truncated to order $\e^3$. The diagonal element is found to be 
\begin{align}
H^{{\rm eff}(2)}_{nn} = H^{{\rm eff}(1)}_{nn} + \e^3 \bigg[-\frac{1}{2} \omega_{\m\n} \dot{x}^{\m} \ddot{x}^{\n} - \frac{1}{6} \gamma_{\lambda\m\n} \dot{x}^{\lambda} \dot{x}^{\m} \dot{x}^{\n} \bigg]\,,
\label{eq:Heff:2}
\end{align}
where
\begin{subequations}  
\label{eq:omega:gamma}
\begin{align}
\label{eq:omega}
\omega_{\m\n} &= -2\mathrm{Im} \langle D_{\m} n|(E_n-H)^{-2}|D_{\n} n\rangle\,, \\[4pt]
\label{eq:gamma}
\gamma_{\l\m\n} &= -6\mathrm{Im} \langle D_{(\l} n|(E_n-H)^{-2} |D_{\m} D_{\n)} n\rangle \nn \\[4pt]
&\quad - 6\mathrm{Im} \langle D_{(\l} n|(E_n-H)^{-1} \partial_{\m} (E_n-H)^{-1} |D_{\n)} n\rangle\,{.}
\end{align}
\end{subequations}
Comparing with the alternative approach based on the approximate solution with eigenstate corrections in Eq.~\eqref{eq:psi p-order}, the third order terms in Eq.~\eqref{eq:Heff:2} are found to be equal to $-\e^3 \dot{\alpha}_2$.  Here $\omega$ is a tensor that induces an almost symplectic structure on the parameter space (nuclear configuration space). Moreover, $\gamma$ is symmetrized with respect to permutations of its three indices.{\footnote{Our convention for symmetrized indices is $a_{(\mu\nu)}=\sfrac 12 (a_{\mu\nu}+a_{\nu\mu})$. Accordingly, for three indices we have $a_{(\mu\nu\rho)}=\sfrac 16(a_{\mu\nu\rho}+a_{\mu\rho\nu}+a_{\rho\mu\nu}+a_{\rho\nu\mu}+a_{\nu\rho\mu}+a_{\nu\mu\rho})$. Antisymmetrization of indices follows suit and is denoted with square brackets.}} 
It is not a tensor but instead obeys the affine transformation law
\begin{align}
\label{eq:transformation law}
\gamma_{\r\m\n} \rightarrow \tilde{\gamma}_{\r'\m'\n'} &= \frac{\p x^{\r}}{\p \tilde{x}^{\r'}} \gamma_{\r\m\n} \frac{\p x^{\m}}{\p \tilde{x}^{\m'}} \frac{\p x^{\n}}{\p \tilde{x}^{\n'}} + 3\frac{\p x^{\r}}{\p \tilde{x}^{(\r'}} \omega_{\r\s} \frac{\p^2 x^{\s}}{\p \tilde{x}^{\m'} \p \tilde{x}^{\n')}}
\end{align}
under the coordinate transformation $x^{\m}\rightarrow \tilde{x}^{\m}=\tilde{x}^{\m}(x)$. The $\gamma$ defined in Eq.~\eqref{eq:gamma}, which we obtained by evaluating the second order effective Hamiltonian in a quantum-classical system, turns out to obey the same transformation law as the fully symmetric object $\gamma$ in Ref.~\cite{Panzhenskii}.  $\gamma$ contains a second covariant derivative $|D_{\m} D_{\n} n\rangle$. To break this down into gauge and coordinate invariant terms, we first decompose it into a horizontal and vertical part with respect to the splitting introduced via the Berry connection: 
\begin{equation}
      D_{\mu}\ket{D_{\nu}n}=Q\ket{D_{\mu}D_{\nu}n}+\ket{n}\braket{n|D_{\mu}D_{\nu}n}\,.
\end{equation} 
This equation holds trivially by the definition of the projector $Q=1-|n\rangle\langle n|$. If we make the additional assumption that the horizontal subspace is generated by $\ket{D_{\mu}n}$, namely that the number of independent parameters $N$ equals the (real) dimension of the projective Hilbert space $2d-2$, we may expand the horizontal part as 
\begin{equation}\label{maximal}
      Q\ket{D_{\mu}D_{\nu}n}=\Upsilon^{\lambda}_{\mu\nu}\ket{D_{\lambda}n}\,,
\end{equation}
for some complex coefficients $\Upsilon^{\lambda}_{\mu\nu}$.
In general, however, this assumption is very strong and it excludes a wide spectrum of interesting cases. Instead we may use a projector $Q'$, defined by
\begin{equation}\label{eq: Q' def}
    Q'= 1-\ket{n}\bra{n}-\ket{D_{\mu}n}h^{\mu\nu}\bra{D_{\nu}n}\,,
\end{equation} 
to project to the subspace that is normal to the direct sum of the vertical ($\ket{n}$) and parameter-generated horizontal (generated by $\ket{D_{\m} n}$) subspaces.  The projector $Q'$ is Hermitian, idempotent and satisfies $Q'\ket{n}=0$ and $Q'\ket{D_{\mu}n}=0$. Note that the inverse of the quantum geometric tensor is used in the definition of $Q'$. 
We can use $Q'$ to decompose $Q\ket{D_{\mu}D_{\nu}n}$ as
\begin{equation}\label{eq:deco Q on second derivative}
Q\ket{D_{\mu}D_{\nu}n}=Q'\ket{D_{\mu}D_{\nu}n}+\ket{D_{\lambda}n}h^{\lambda\kappa}\braket{D_{\kappa}n|D_{\mu}D_{\nu}n}\,.    
\end{equation}
Let us define the quantities $\ket{\Sigma_{\mu\nu}}:=Q'\ket{D_{\mu}D_{\nu}n}$,
which generate a subspace of the horizontal subspace that is normal to all $\ket{D_{\mu}n}$, i.e. they satisfy 
$\braket{D_{\lambda}n|\Sigma_{\mu\nu}}=0$.
In this generic decomposition, we have identified 
\begin{equation}
\Upsilon^{\lambda}_{\mu\nu}=h^{\lambda\kappa}\braket{D_{\kappa}n|D_{\mu}D_{\nu}n}\,.
\end{equation}
The coefficients $\Upsilon^{\lambda}_{\mu\nu}$ have a geometrical meaning. They are the components of a covariant derivative for quantum tangent vectors of the form $\ket{V}=V^{\mu}\ket{D_{\mu}n}$, in other words they tells us how to parallel transport such vectors along the parameter manifold \cite{Requist:2022vox}. Denoting this quantum covariant derivative by $\widehat{\nabla}$, we have 
\begin{equation}
      \widehat{\nabla}_{X}\ket{V}=X^{\mu}(\partial_{\mu}V^{\lambda}+\Upsilon^{\lambda}_{\mu\nu}V^{\nu})\ket{D_{\lambda} n}\,.
\end{equation}
The torsion of this connection vanishes as a consequence of the orthogonality condition \eqref{eq: orthogonality}, namely the connection is symmetric: $\Upsilon^{\lambda}_{\mu\nu}=\Upsilon^{\lambda}_{\nu\mu}$. Additionally, a direct computation shows that it is compatible with the quantum geometric tensor, $\widehat{\nabla}h=0$, which reads
\begin{align}
    \partial_{\lambda}h_{\mu\nu}=\bar{\Upsilon}^{\rho}_{\lambda\mu}h_{\rho\nu}+\Upsilon^{\rho}_{\lambda\nu}h_{\mu\rho}\,.
\end{align}
These two conditions alone do not completely determine the real and imaginary components of the connection. The real part of the metricity condition leads to 
\begin{align}
    \mathrm{Re}(\Upsilon^{\lambda}_{\mu\nu})=\mathring\Gamma[g]^{\lambda}_{\mu\nu}- J^{\lambda}{}_{\kappa}\,\mathrm{Im}(\Upsilon^{\kappa}_{\mu\nu})\,,
\end{align}
expressing the real part of the connection coefficients in terms of the Levi-Civita connection of the Riemannian metric $g$ (the real part of the quantum geometric tensor), the imaginary part of the connection coefficients and the $(1,1)$-tensor $J=\sfrac 12 g^{-1}\circ B: T\mc{M}\to T\mc{M}$, an endomorphism of the tangent bundle, whose components are 
\begin{align}
    J^{\mu}{}_{\nu}=\sfrac 12 \, g^{\mu\kappa}B_{\kappa\nu}=-J_{\nu}{}^{\mu}\,,
\end{align}
the latter being the transpose $J^{T}=\sfrac 12 B\circ g^{-1}: T^{\ast}\mc{M}\to T^{\ast}\mc{M}$. In the maximal case, i.e.~when $N=2d-2$ so that the parameter space inherits the K\"ahler structure of the projective Hilbert space, the endomorphism $J$ is a complex structure, namely $J^{2}=-1$. This is, however, not true in general. On the other hand, the imaginary part of the metricity condition does not fully determine the imaginary part of the connection coefficients. Instead it constrains them according to 
\begin{align}
    2\,(g_{\rho[\mu}+\sfrac 14 J_{\rho}{}^{\kappa}B_{\kappa[\mu})\,\mathrm{Im}(\Upsilon^{\rho}_{\nu]\lambda})= - \sfrac 12\mathring\nabla_{\lambda}B_{\mu\nu}\,,
\end{align}
where $\mathring\nabla=\mathring\nabla[g]$. 
We may also use the quantum geometric tensor to define the components 
\begin{equation}
      \Upsilon_{\lambda\mu\nu}:=h_{\lambda\kappa}\Upsilon^{\kappa}_{\mu\nu}= \braket{D_{\lambda}n|D_{\mu}D_{\nu}n}\,.
\end{equation}
Then it turns out that 
\begin{equation}    \mathrm{Re}(\Upsilon_{\lambda\mu\nu})=g_{\lambda\sigma}\mathring{\Gamma}^{\sigma}_{\mu\nu}[g]\,,
\end{equation} 
while their imaginary part is related to the Berry curvature according to
\begin{align}
\mathrm{Im}(\Upsilon_{\m\n\r}) - \mathrm{Im} (\Upsilon_{\n\m\r}) &= -\sfrac{1}{2} \partial_{\r} B_{\m\n}\,{.}
\end{align}
This fixes its mixed-symmetry part and leaves its fully symmetric part as an undetermined component.
With the help of these quantities, $\gamma$ can be expressed as
\begin{align}
\label{eq:gamma:alt}
\gamma_{\l\m\n} &= -6\mathrm{Im} \Big[ \langle D_{(\l} n|(E_n-H)^{-2} |D_{\r} n\rangle \Upsilon^{\r}_{\m\n)} \Big] \nn \\
&\quad -6\mathrm{Im} \langle D_{(\l} n|(E_n-H)^{-2}|\Sigma_{\m\n)}\rangle \nn \\[4pt]
&\quad - 6\mathrm{Im} \langle D_{(\l} n|(E_n-H)^{-1} \partial_{\m} (E_n-H)^{-1} |D_{\n)} n\rangle {.}
\end{align}
In the following section, we will show how $\omega$ and $\gamma$ together define a geometric structure on parameter space.

The generators of the near-identity transformations $U_k=e^{i\e^k G_k}$
are constructed in terms of the zeroth order adiabatic eigenstates and eigenvalues. 
Having constructed the generators $G_1$ and $G_2$ with non-vanishing matrix elements 
\begin{align}
&G_{1,ln} = \frac{\langle l| D_\mu n\rangle}{E_l-E_n}\dot{x}^\mu\quad \mathrm{for} \; l\neq n\, ~,\nn \\[4pt]
&G_{2,ln}= \big(\sfrac{i}{(E_n-E_l)^2}\langle l|D_\mu D_\nu n\rangle+ \sfrac{i}{(E_n-E_l)^2}\langle D_\mu l|D_\nu n\rangle-\sfrac{i\partial_\mu(E_n-E_l)}{(E_n-E_l)^3}\langle l|D_\nu n\rangle\nn \\[4pt]
&\hspace{1cm} -\sfrac{i}{2(E_n-E_l)}\langle D_\mu l|(E_l-H)^{-1}|D_\nu n\rangle
-\sfrac{i}{2(E_n-E_l)}\langle D_\mu l|(E_n-H)^{-1}|D_\nu n\rangle \nn \\[4pt]
&\hspace{1cm} +\sfrac{1}{(E_n-E_l)^2}\langle l|D_\nu n\rangle(A_{\mu,ll}-A_{\mu,nn})\big)\dot{x}^\mu\dot{x}^\nu +\sfrac{i}{(E_n-E_l)^2}\langle l|D_\mu n\rangle\ddot{x}^\mu\quad \mathrm{for} \; l\neq n\,~,
\end{align}
we are able to evaluate electronic observables to order $\e^2$ at any point along the nuclear trajectory.  
The average of an electronic observable $\mathcal{O}$ can be calculated 
by transforming to the second order adiabatic representation according to
\begin{align}
    \mathcal{O}^{(2)} =& U^{\dag}_2 U^{\dag}_1 \mathcal{O} U_1 U_2 
\end{align}
and evaluating the expectation value
\begin{align}
    \langle \mathcal{O}\rangle(t)  &= \langle \psi^{{\rm eff}(2)} | \mathcal{O}^{(2)} | \psi^{{\rm eff}(2)} \rangle\big|_{\rm second\, order\, terms} \nn \\
    &= \mathcal{O}_{nn} -i \e [G_1,\mathcal{O}]_{nn} + \frac{1}{2}(-i\e)^2 [G_1,[G_1,\mathcal{O}]]_{nn} -i\e^2 [G_2,\mathcal{O}]_{nn}
    \,.
\end{align}
In this expression all operators are evaluated in the zeroth order adiabatic representation.

The iterative unitary transformation approach employed here is similar to the iterative approach introduced by Berry in Ref.~\cite{berry1987}, and in fact the nonadiabatic induced mass term can be obtained at the first order in Berry's approach \cite{requist2010}.  However, it is difficult to perform analytical calculations to higher order because the $p$th order Hamiltonian is constructed in terms of $p$th order adiabatic eigenstates, which are defined by diagonalizing the $(p-1)$th order Hamiltonian to all orders in $\e$. 

The type of adiabatic perturbation theory developed here is conceptually similar to Lie-transform perturbation theory in classical mechanics. Our sequence of near-identity unitary transformations, which generate a hierarchy of adiabatic representations, is a quantum mechanical counterpart of the sequence of near-identity canonical transformations in Lie-transform perturbation theory
\cite{hori1966,delprit1969,dragt1976,cary1981}. For two-component systems in which both components are treated quantum mechanically, adiabatic perturbation theory has been developed using the Wigner-Weyl-Moyal formalism with a convention for ordering the perturbations that is suitable for typical molecular states \cite{littlejohn1993,weigert1993,panati2003,matyus2019,littlejohn2024} and using near-identity unitary transformations with a different ordering convention \cite{requist2025}.

\section{Effective actions and equations of motion}
\label{sec:EffEOM}

To obtain a hierarchy of increasingly accurate effective equations of motion for $x$, we seek a hierarchy of effective actions. 
For this purpose, we need to approximate the electronic part 
\begin{align}
\label{eq:electronic action}
S_{elec}[x,\ket{\psi}] = \int \dd t \,\langle \psi | i\e \frac{d}{dt} - H(x)|\psi\rangle
\end{align}
of the action in Eq.~\eqref{eq:bare action} by a $p$th order effective action $S_{elec}^{{\rm eff}(p)}[x]$ that is a functional of the trajectory $x$ but not the state $\ket{\psi}$.

The hierarchy of effective Hamiltonians derived in the previous section leads directly to a hierarchy of effective actions. Writing Eq.~\eqref{eq:electronic action} in the $p$th order adiabatic representation and approximating $H^{(p)}$ by $H^{{\rm eff}(p)}$ and $|\psi^{(p)}\rangle$ by
$|\psi^{{\rm eff}(p)}\rangle$
yields the effective action
\begin{align}
\label{eq:electronic action:eff}
S_{elec}^{{\rm eff}(p+1)}[x] = \int \dd t\, \langle \psi^{{\rm eff}(p)}[x] | i\e \frac{d}{dt} - H^{{\rm eff}(p)}(x,\dot{x},\ldots,x^{(p+1)})|\psi^{{\rm eff}(p)}[x]\rangle\,{,}
\end{align}
which is a functional of $x$ because $|\psi^{{\rm eff}(p)}[x]\rangle$ is a functional of $x$.  The notation $x^{(k)}$ denotes the $k$th order time derivative of $x$. Consider the variation with respect to $x$. Since  
\begin{align}
|\psi^{{\rm eff}(p)}[x]\rangle = e^{-i\e^{-1}\int_{t_0}^t \dd s \, H^{{\rm eff}(p)}_{nn}(s)} |n^{(p)}\rangle 
\end{align}
is the solution of the $p$th order effective Schr\"odinger equation with the initial condition $\ket{\psi^{(p)}(t_0)}=\ket{n^{(p)}(x(t_0))}$, 
there is no contribution from the variation with respect to the wave function (via the chain rule) and hence the variation of $S_{elec}^{{\rm eff}(p+1)}$ is the same as the variation of the action
\begin{align}
\label{eq:electronic action:eff:alt}
S_{elec}^{{\rm eff}(p+1)\prime}[x] = -\int \dd t \, H^{{\rm eff}(p)}_{nn}(x,\dot{x},\ldots,x^{(p+1)}){.}
\end{align}
This establishes that the $(p+1)$th order effective electronic action $S_{elec}^{{\rm eff}(p+1)\prime}$ is precisely $\e$ times the time-dependent phase of $\ket{\psi^{{\rm eff}(p)}}$. 
Importantly, the accuracy of the effective action is one order higher than the order of the adiabatic representation in which the effective Hamiltonian is expressed.  
Since $|\psi^{{\rm approx},p}\rangle$ has the same time-dependent phase factor as $|\psi^{{\rm eff}(p)}\rangle$, the nonintegrable phase $\alpha_p$ provides the $(p+1)$th order term in the effective action.

We will now present the hierarchy of effective actions up to third order, discussing their building blocks as a way of highlighting the subtleties of the third order.

\paragraph{First order effective action.}  
The building blocks of the first order Lagrangian are well-known.
They are the classical kinetic energy inherited from the original action in Eq.~\eqref{eq:bare action}, a potential $E_n$ that produces the Born-Oppenheimer force, and a velocity-dependent gauge coupling that produces an effective Lorentz force. 
Using the diagonal element of $H^{{\rm eff}(0)}$ in Eq.~\eqref{eq:Heff0}, the first order effective action is found to be
\begin{equation}
    S^{{\rm eff}(1)} = \int\dd t \bigg[\frac 12 M_{\mu\nu}(x)\,\dot{x}^{\mu}\dot{x}^{\nu}-E_n+\epsilon A_{\mu}\dot{x}^{\mu}\bigg]\,.
\end{equation}
The Euler-Lagrange equations are 
\begin{equation}
\label{eq:EL:1}
    M_{\mu\nu}{a}^{\nu}-\epsilon B_{\mu\nu}\dot{x}^{\nu}+\partial_{\mu}E_n=0\,,
\end{equation}
where we recall that $a^{\mu}$ is the acceleration vector with respect to the Levi-Civita connection for the bare mass tensor $M_{\mu\nu}$ and $B_{\mu\nu}$ is the Berry curvature. Exactly the same equation is found by using the first order adiabatic solution $|\psi^{{\rm approx},1}\rangle$ to evaluate the effective force $-\langle \psi|\partial_{\m}H|\psi\rangle$ in Eq.~\eqref{eq:Newton law:scaled time}.
The Berry-Lorentz force is important in molecular dynamics simulations in an external magnetic field \cite{ceresoli2007,culpitt2021} and in the presence of electronic currents in nanoscale conductors \cite{lu2010,todorov2014}.

\paragraph{Second order effective action.} 

The only new term in the second order Lagrangian comes from Eq.~\eqref{eq:second-order:terms} and contains the nonadiabatic mass correction $M_{2\m\n}$.
The second order effective action is 
\begin{equation}
\label{eq:action:2ndorder}
    S^{{\rm eff}(2)}=S^{{\rm eff}(1)} + \e^2\int \dd t \,\bigg[\frac{1}{2} M_{2\m\n}(x) \dot{x}^{\m} \dot{x}^{\n}\bigg]\,.    
\end{equation}
The Euler-Lagrange equations are
\begin{equation}
\label{eq:EL:2}
    M^{\star}_{\mu\nu}{a}^{\star\nu}-\epsilon B_{\mu\nu}\dot{x}^{\nu}+\partial_{\mu}E_n=0\,,
\end{equation}
where we defined
\begin{equation} M^{\star}_{\m\n}=M_{\m\n}+\e^2 M_{2\m\n}\,,
\end{equation} 
the renormalized mass tensor, and $a^{\star}$ is the acceleration vector with respect to the Levi-Civita connection $\mathring\nabla[M^{\star}]$  for $M^{\star}$.{\footnote{This is based on the fact that given two Riemannian metrics, say $g_1$ and $g_2$, with corresponding Levi-Civita connections $\mathring\nabla^1\equiv \mathring\nabla[g_{1}]$ and $\mathring\nabla^2\equiv \mathring\nabla[g_2]$, the sum of them is a Riemannian metric $g=g_1+g_2$ with Levi-Civita connection $\mathring\nabla\equiv\mathring\nabla[g]$ which for any vector fields $X, Y, Z$ satisfies
\begin{equation}\label{eq:sum of metrics}
    g(\mathring\nabla_{X}Y,Z)=g_1(\mathring\nabla^1_XY,Z)+g_2(\mathring\nabla^{2}_{X}Y,Z)\,.
\end{equation}
Applying \eqref{eq:sum of metrics} for both $X, Y$ being the velocity vector and for $g_1=M$ and $g_2=\epsilon^2 M_2$, we obtain
\begin{equation}
    M^{\star}_{\mu\nu}\,a^{\star\,\nu}=M_{\mu\nu}\,a^{\nu}+ \,\epsilon^2M_{2\mu\nu}\,\mathring{a}_2^{\nu}\,,
\end{equation}
where $\mathring{a}_{2}$ is the acceleration vector with respect to $\mathring\nabla[M_2]$, which proves \eqref{eq:EL:2}.}}
The system of second order (Newtonian) differential equations in Eq.~\eqref{eq:EL:2} have the same form as Eq.~\eqref{eq:EL:1} apart from a renormalization of the mass tensor and acceleration.
Eq.~\eqref{eq:EL:2} is equivalent to the equation of motion obtained by evaluating the effective force to second order.
In terms of the momenta $p_{\m} = M^{\star}_{\m\n} \dot{x}^{\n} + A_{\m}$, the energy to second order in $\e$ is
\begin{align}
\mathcal{H}^{{\rm nucl}(2)} 
&= \frac{1}{2} M^{\star\m\n} (p_{\m}-A_{\m})(p_{\n}-A_{\n}) + E_n\,. 
\end{align}
It is worth noting the following distinction between the electronic and nuclear effective Hamiltonians: 
whereas the $M_{2\m\n}\dot{x}^{\m}\dot{x}^{\n}$ term enters the nuclear Hamiltonian $\mathcal{H}^{{\rm nucl}(2)}$ with a positive sign, it enters the electronic Hamiltonian with a negative sign.

\paragraph{Third order effective action.} 
The least straightforward and most interesting part of the problem appears at third order.  The third order effective action is
\begin{equation}
\label{eq:action:3rdorder}
    S^{{\rm eff}(3)}=S^{{\rm eff}(2)} + \e^3\int \dd t \,\bigg[\frac{1}{2} \omega_{\m\n}(x) \dot{x}^{\m} \ddot{x}^{\n} +\frac{1}{6} \gamma_{\l\m\n}(x) \dot{x}^{\l}\dot{x}^{\m}\dot{x}^{\n}\bigg] \,.    
\end{equation}
To determine the effective equations of motion, recall that for an acceleration-dependent Lagrangian density $L(x^{\mu},\dot{x}^{\mu},\ddot{x}^{\mu},t)$, the Euler-Lagrange equations read \cite{Borneas}
\begin{equation}\label{eq:EL higher general formula}
   \frac{\partial L}{\partial x^{\mu}}-\frac{\dd}{\dd t}\frac{\partial L}{\partial\dot{x}^{\mu}}+\frac{\dd^2}{\dd t^2}\frac{\partial L}{\partial \ddot{x}^{\mu}}=0\,.
\end{equation}
In the present case, the Euler-Lagrange equations are
\begin{equation}
\label{eq:EL:3}
    \e^3 \omega_{\mu\nu}\,\dddot{x}^{\nu}+ \e^3 \zeta_{\mu\kappa\lambda}\,\dot{x}^{\kappa}\ddot{x}^{\lambda}+\e^3 \xi_{\mu\kappa\lambda\rho}\,\dot{x}^{\kappa}\dot{x}^{\lambda}\dot{x}^{\rho} + M^{\star}_{\mu\nu}{a}^{\star\nu}-\epsilon B_{\mu\nu}\dot{x}^{\nu}+\partial_{\mu}E_n=0\,,
\end{equation}
where
\begin{subequations}
\label{eq:zeta:xi}
\begin{align}
\label{eq:zeta}    \zeta_{\mu\kappa\lambda}&=-\sfrac 12 H_{\mu\kappa\lambda}-2\partial_{(\kappa} \omega_{\lambda)\mu}+\gamma_{\mu\kappa\lambda}\,, \\[4pt]
\label{eq:xi}
    \xi_{\mu\kappa\lambda\rho}&=  -\sfrac 16 (\partial_{\mu}\gamma_{\kappa\lambda\rho}-3\,\partial_{(\rho}\gamma_{\kappa\lambda)\mu}+3\,\partial_{(\kappa}\partial_{\lambda}\omega_{\rho)\mu})\,.
\end{align} 
\end{subequations}
Here the fully antisymmetric 3-tensor $H_{\mu\kappa\lambda}$ is defined as 
\begin{align}
    H_{\mu\kappa\lambda}=\partial_{\mu}\omega_{\kappa\lambda}+\partial_{\lambda}\omega_{\mu\kappa}+\partial_{\kappa}\omega_{\lambda\mu}\,;
\end{align}
in differential form notation it is the exterior derivative of the almost symplectic 2-form, namely $H=\dd\omega$. 
We have decomposed $\zeta$ into fully antisymmetric, fully symmetric, and mixed-symmetric parts such that the former is given precisely by minus half of $H$, the latter via a symmetrized derivative of $\omega$, and the fully symmetric part by the quantity $\gamma$ in Eq.~\eqref{eq:gamma}.  The quantity $\xi$ is fully symmetric with respect to its last three indices.  

The equation of motion \eqref{eq:EL:3} is form-invariant under a coordinate transformation, although this is not obvious for the third order terms.  For instance, $\zeta$ and $\xi$ are not tensors. In the Section \ref{sec:jerk}, we will put the equation of motion in a manifestly covariant form to bring to light the geometrical significance of the third order terms.

The equation of motion \eqref{eq:EL:3} is non-Newtonian due to the presence of terms that are third order in time derivatives.  We report, moreover, that in the alternative approach described around Eq.\,\eqref{eq:psi p-order} evaluating the effective force to third order in $\e$ leads to an equation of motion that is identical to Eq.~\eqref{eq:EL:3} except for the addition of the following third order terms:
\begin{align}
\label{eq:BeeGee}
-\frac{1}{2} \Big[ B_{\m(\l} g_{3\k\r)} +\omega_{\m(\l} g_{\k\r)}\Big] \dot{x}^{\l} \dot{x}^{\k} \dot{x}^{\r} \,,
\end{align}
where $g_{3}$ is a symmetric tensor which is obtained as the real part of a higher analog of the quantum geometric tensor
\begin{equation}
h_{3\mu\nu}:=\bra{D_{\mu}n}(E_n-H)^{-2}\ket{D_{\nu}n}=g_{3\mu\nu}-\frac i2\omega_{\mu\nu}\,,
\end{equation} 
in agreement with the identification of $\omega_{\mu\nu}$ in Eq.~\eqref{eq:omega}.
Interestingly, the terms in Eq.~\eqref{eq:BeeGee} are non-Lagrangian. Nevertheless, they do not spoil the conservation of energy associated with Eq.~\eqref{eq:EL:3}, since they do no work, i.e.~the product of Eq.~\eqref{eq:BeeGee} with $\dot{x}^{\m}$ is identically zero due to the antisymmetry of the $B$ and $\omega$ factors.

For a Lagrangian that depends on second order time derivatives, the energy is \cite{Borneas}
\begin{align}
\mathcal{E} &= 
\left[\frac{\partial L}{\partial \dot{x}^{\m}} - \frac{d}{dt} \left( \frac{\partial L}{\partial \ddot{x}^{\m}} \right)  \right]\dot{x}^{\m} + \frac{\partial L}{\partial \ddot{x}^{\m}} \ddot{x}^{\m} - L\,.
\label{eq:H:canonical}
\end{align}
Substituting the effective Lagrangian $L^{{\rm eff}(3)}$ corresponding to the action in Eq.~\eqref{eq:action:3rdorder} into Eq.~\eqref{eq:H:canonical} gives
\begin{align}
\mathcal{E}^{{\rm nucl}(3)} = \e^3 \left[ \omega_{\m\n} \dot{x}^{\m} \ddot{x}^{\n} + \frac{1}{3} \gamma_{\l\m\n} \dot{x}^{\l} \dot{x}^{\m} \dot{x}^{\n} \right] + \frac{1}{2} M^{\star}_{\m\n} \dot{x}^{\m} \dot{x}^{\n} + E_n \,.
\end{align}
With respect to $\mathcal{H}^{{\rm nucl}(2)}$, the energy acquires the third order term
\begin{align}
\e^3 \omega(v,\mathring{a}) \,,
\end{align}
where $\mathring{a}$ is the covariant acceleration with respect to a symmetric connection that arises at third order and will be defined explicitly in the following section.
Since $L^{{\rm eff}(3)}$ has no explicit time dependence, $\mathcal{E}^{{\rm nucl}(3)}$ is conserved by the third order equation of motion.  A third order Hamiltonian can be defined as a function of Ostrogradsky's generalized coordinates and momenta.

\section{Effective Kawaguchi geometry}

\subsection{Kawaguchi manifolds of order 2}  

A remarkable conclusion drawn from the previous section is that the effective action of the mixed quantum-classical system at third order in the adiabatic parameter is a simple action functional that depends on the second order time derivative of the nuclear positions. Unlike the effective action up to second order, which is based on Riemannian geometry, to understand the geometrical underpinnings of the third order effective action we must appeal to a different kind of geometry. 

A useful way to describe this geometry is the following. First, let us recall the statement of S.\,S.\,Chern that \emph{Finsler geometry is not a generalization of Riemannian geometry. It is better described as Riemannian geometry without the quadratic restriction} \cite{Chern}. Given a smooth manifold $\mc{M}$, a Finsler structure is a function 
\begin{equation}
    F: T\mc{M} \to [0,\infty)\,,
\end{equation}
which satisfies the following properties: (i) Regularity: it is smooth on the tangent bundle with the zero section removed (the slit tangent bundle $T_0\mc{M}=T\mc{M}\smallsetminus \{0\}$), (ii) Positive homogeneity: $F(x,\lambda y)=\lambda F(x,y)$ for all positive $\lambda$, where $x^{\mu}$ are the base coordinates and $y^{\mu}$ are the fiber coordinates of the tangent bundle, and (iii) Strong convexity: the Hessian matrix 
\begin{equation}
    g_{\mu\nu}=\frac 12 \frac{\partial^{2}F^{2}}{\partial y^{\mu}\partial y^{\nu}}\,,
\end{equation}
is positive-definite at every point of $T_{0}\mc{M}$. A Finsler manifold $(\mc{M},F)$ is a pair of a smooth manifold and a Finsler structure on it \cite{Bao}. Note that for particular applications sometimes regularity and strong convexity are too strong and they can be relaxed \cite{Tanaka:2013osa}.

The quadratic restriction is that the Hessian is a function of the base coordinates only, in other words that the Finsler metric is quadratic in the fiber coordinates. In that case we end up with a Riemannian manifold, for which 
\begin{equation}
    F(x,y)=\sqrt{g_{x}(y,y)}\,,
\end{equation}
where $g_{x}$ is a family of inner products for each tangent space of the manifold $\mc{M}$.

To account for higher derivatives, the domain of the function $F$ has to be replaced with a higher order tangent bundle. This was studied already in the 1930s \cite{Kawaguchi} and goes under the name of Kawaguchi geometry. We refer to the PhD thesis \cite{Tanaka:2013osa} for a modern treatment of the subject and we discuss here only the basic elements that we need for our purposes. Since the total space of the tangent bundle over a smooth manifold is itself a smooth manifold, it is possible to construct its tangent bundle. This construction results in the second order tangent bundle $T(T\mc{M})\equiv T^{2}\mc{M}$, also known as the double tangent bundle, see e.g.~Ref.~\cite{Mackenzie_2005}. The slit double tangent bundle $T^2_0\mc{M}$ is obtained by removing the zero section. Moreover, consider coordinates $(x^{\mu},y^{\mu},z^{\mu})$ in a local chart, $z^{\mu}$ being fiber coordinates of second order. We emphasize that $T^{2}\mc{M}$ is a vector bundle over the total space $T\mc{M}$, but it is \emph{not} a vector bundle over the base manifold $\mc{M}$.{\footnote{It is, however, possible to turn it into a vector bundle over $\mc{M}$ at the expense of additional, noncanonical data. Specifically, a choice of connection $\nabla$ on $T\mc{M}$ splits (linearizes) it to the direct sum $T^{2}\mc{M}\simeq T\mc{M}\oplus T\mc{M}$, and the right-hand side is a vector bundle over $\mc{M}$. We will use this later in our discussion.\label{footnote}} We can now define a Kawaguchi structure of order 2 as a function 
\begin{equation}
    K: T^{2}\mc{M}\to [0,\infty)\,,
\end{equation}
which satisfies (i) Regularity: it is smooth on the slit double tangent bundle, and (ii) Positive second order homogeneity: for all positive $\lambda$, the function satisfies 
\begin{equation}
    K(x,\lambda y, \lambda^2 z+\mu y)=\lambda K(x,y,z)\,, \quad \mu\in\R\,.
\end{equation}
A Kawaguchi manifold $(\mc{M},K)$ of order 2 is then a smooth manifold $\mc{M}$ endowed with a Kawaguchi structure of order 2.  

The following remarks are in order. There is no convexity property in general, since this would exclude the physical examples we are interested in. Moreover, the structure can be defined accordingly for any positive integer; however, for the purposes of this paper, order 2 is sufficient and we focus on it---higher order Kawaguchi structures would be relevant at higher orders in adiabatic perturbation theory. 

The main motivation behind this definition is to be able to define a notion of length for a curve which is given by a reparametrization invariant integral. Indeed, the definition is tailored so that the higher order generalization of the Euler theorem, the Zermelo conditions 
\begin{subequations} 
\begin{align}
    y^{\mu}\frac{\partial K}{\partial y^{\mu}}+2z^{\mu}\frac{\partial K}{\partial z^{\mu}}=K\,, \qquad y^{\mu}\frac{\partial K}{\partial z^{\mu}}=0\,,
\end{align}
\end{subequations}
originally stated in the PhD thesis of E.\,Zermelo \cite{Zermelo}, are satisfied \cite{Tanaka:2013osa,Urban}. The reparametrization invariant integral, the ``arc length'' of a curve in Kawaguchi geometry, is given as 
\begin{equation}
    S_{\text{K}}=\int_{t_i}^{t_{f}} \dd t \, K\left(x^{\mu},\dot{x}^{\mu},\ddot{x}^{\mu}\right)\,.
\end{equation}

The general form of this arc length should be specialized to match the effective action we found in the previous section. The next step in this task is to define a special kind of Kawaguchi geometry that would accommodate the geometric structures $\omega_{\mu\nu}$ and $\gamma_{\mu\nu\rho}$ that emerged from the effective Hamiltonian at third order. This is achieved through the following quadratic and cubic constraints on the fundamental function $K$: 
\begin{subequations}\label{eq: quadratic and cubic restrictions}
    \begin{align}
 &   \omega_{\mu\nu}:=\frac{\partial^{2}K^3}{\partial y^{\mu}\partial z^{\nu}}=\omega_{\mu\nu}(x)\,,\\[4pt]
  & \gamma_{\mu\nu\rho}:=\frac 16\frac{\partial^3K^3}{\partial y^{\mu}\partial y^{\nu}\partial y^{\rho}}= \gamma_{\mu\nu\rho}(x)\,\\[4pt] 
  & \chi_{\mu\nu}:=\frac 12 \frac{\partial^2 K^3}{\partial z^{\mu}\partial z^{\nu}}=0\,,
\end{align}
\end{subequations}
together with the condition that $\omega_{\mu\nu}$ is non-degenerate. Due to the third condition, the system is singular from a general perspective. This is expected in view of the fact that $\ddot{x}$-dependent Lagrangians generically lead to fourth order equations of motion, whereas for the purposes of the present work we would like to obtain third order equations.

Due to the homogeneity property of $K$, $\omega_{\mu\nu}$ is a skew-symmetric $(2,0)$ tensor (a 2-form) and $\gamma_{\mu\nu\rho}$ is fully symmetric and it transforms inhomogeneously as in Eq.~\eqref{eq:transformation law}. A non-degenerate 2-form is an almost symplectic structure and it only exists in even dimensions. This special class of Kawaguchi manifolds of order 2 was found by Losik in Ref.~\cite{Losik}, called $L^2$ there, and were dubbed Kawaguchi manifolds with a special metric in Ref.~\cite{Panzhenskii}. They were also described earlier by Lemleyn \cite{Lemleyn} without a direct connection to Kawaguchi manifolds. Hence we refer to them as Kawaguchi-Lemleyn-Losik manifolds of order 2 or KL$^2$ manifolds. The construction resembles our earlier discussion on the relation between Riemannian and Finsler manifolds; KL$^2$ manifolds play the same role as Riemannian manifolds, but for Kawaguchi manifolds of order 2.

As with Riemannian manifolds, which can be defined without direct reference to the Finsler function $F$, KL$^2$ manifolds can be defined without reference to $K$ \cite{Lemleyn}. Recall that a nondegenerate 2-form $\omega\in\Omega^{2}(\mc{M})$ on a manifold is an almost symplectic structure. It becomes symplectic when it is closed, namely when $\dd\omega=0$ under the action of the de Rham differential. We do not require closure, however. Consider a symmetric affine connection $\mathring\nabla$ on the almost symplectic manifold and impose the condition that it is proportional to $\dd\omega=H$, in particular that 
\begin{equation}
    \label{eq:almost compatible connection}
\mathring\nabla_{\mu}\omega_{\nu\rho}=\sfrac 13 H_{\mu\nu\rho}\,.
\end{equation}
We note that a symmetric connection cannot be compatible with an almost symplectic structure; compatible connections necessarily have torsion. On the other hand, in the symplectic case (when $H=0$,) a symmetric connection $\mathring\nabla$ satisfying this compatibility condition, namely $\mathring\nabla\omega=0$, is called a symplectic connection. Recall that in Riemannian geometry a symmetric connection compatible with the metric is unique, the Levi-Civita connection. On the other hand, in symplectic geometry, this is no longer the case. A symmetric connection $\mathring\nabla$ compatible with the symplectic form always exists, however it is not unique. The space of such symplectic connections is an affine space modelled on the space of contravariant fully symmetric 3-tensors \cite{bieliavsky2006symplecticconnections}. Similarly, in the almost symplectic case, a connection that satisfies Eq.~\eqref{eq:almost compatible connection} is not unique. However, any such connection gives rise to a geometric object $\gamma_{\mu\nu\rho}$ that transforms according to Eq.~\eqref{eq:transformation law} by means of 
\begin{equation}\label{eq:def of gamma}
    \sfrac 13 \gamma_{\mu\nu\rho}:=\omega_{\nu\sigma}\mathring{\Gamma}^{\sigma}_{\mu\rho}+\sfrac 23 \partial_{(\mu}\omega_{\rho)\nu}\,,
\end{equation}
where $\mathring\Gamma^{\mu}_{\nu\rho}$ are the components of $\mathring\nabla$ in a coordinate chart. This $\gamma_{\mu\nu\rho}$ is fully symmetric, but it does not transform tensorially. A tensorial object, namely a fully symmetric covariant 3-tensor $\Phi$ may be defined as 
\begin{equation}
    \Phi_{\mu\nu\rho}:=\gamma_{\mu\nu\rho}+3\,\omega_{\sigma(\mu}\mathring{\Gamma}'^{\sigma}_{\nu\rho)}\,,
\end{equation}
for any symmetric connection $\mathring\nabla'$ satisfying Eq.~\eqref{eq:almost compatible connection}. This tensor $\Phi$ generates the family of such connections and the choice Eq.~\eqref{eq:def of gamma}, namely $\mathring\nabla'=\mathring\nabla$, corresponds to $\Phi=0$. In Ref.~\cite{Lemleyn}, an almost symplectic manifold with a structural object $\gamma_{\mu\nu\rho}$ is called a ``space of symmetric almost symplectic connection''. In Ref.~\cite{Losik}, a Kawaguchi manifold in the class $L^{2}$ is shown to be a space of symmetric affine connection with the covariant 2- and 3-tensors $\omega$ and $\Phi$. 

According to these, the arc length we consider is given with respect to a fundamental function $K$ that has the form of a cubic root and involves the geometric objects $\omega_{\mu\nu}$ and $\gamma_{\mu\nu\rho}$:
  \begin{equation}
    S_{\text{KL$^2$}}=\int \dd t \, \sqrt[3]{\sfrac 12\, \omega_{\mu\nu}(x)\dot{x}^{\mu}\ddot{x}^{\nu}+\sfrac 16\,\gamma_{\mu\nu\rho}(x)\dot{x}^{\mu}\dot{x}^{\nu}\dot{x}^{\rho}}\,.\label{eq:cubic root action}
\end{equation}
The action in Eq.~\eqref{eq:cubic root action} describes propagation in a KL$^2$ manifold, where the generalized arc length of a curve is given by that integral. Since the 2-form $\omega$ is nondegenerate, we can rewrite this action in the completely equivalent form
\begin{equation}
    S_{\text{KL$^2$}}=\int \dd t \, \sqrt[3]{\sfrac 12\,\omega_{\mu\nu}v^{\mu}\mathring{a}^{\nu}}\,,
\end{equation}
where $v^{\mu}=\dot{x}^{\mu}$ is the velocity and $\mathring{a}^{\mu}$ are the components of a covariant acceleration vector defined through the symmetric connection that satisfies Eq.~\eqref{eq:almost compatible connection} and generates $\gamma_{\mu\nu\rho}$ through Eq.~\eqref{eq:def of gamma} (the $\Phi=0$ case.) The fact that the action is now written in terms of proper vectors is due to the connection that split the second order bundle $T^2\mc{M}$ into a direct sum bundle over the base $\mc M$, cf. footnote \ref{footnote}.

Now that we have set up the structures underlying the effective geometry for our problem, we are almost in a position to explain how they match with the third order effective action. We observe, however, that the action of the Kawaguchi particle just described involves a cubic root, unlike the effective action Eq.~\eqref{eq:action:3rdorder}. Associated to this is the fact that although the action $S_{\text{KL$^2$}}$ is by construction reparametrization invariant, the action Eq.~\eqref{eq:action:3rdorder} does not manifest this symmetry. To explain this discrepancy and show that the two actions are in fact classically equivalent Lagrangian systems, let us first recall the analogous situation for the ordinary relativistic particle. 

\subsection{A reminder: the relativistic particle} \label{subsec:relativistic particle}

The material presented here is standard and may be found in any textbook on string theory. Here we mainly follow Ref.~\cite{BLT}. Consider a free relativistic particle with mass $m$, propagating in $D$-dimensional Minkowski spacetime with metric $\eta_{\mu\nu}$. The action functional is given by the length of the worldline which is swept by the point particle during its motion, 
\begin{equation}
    S=-m\int_{s_i}^{s_f}\dd x = -m\int_{\tau_i}^{\tau_f}\dd\tau \sqrt{-\eta_{\mu\nu}\dot{x}^{\mu}\dot{x}^{\nu}}=-m\int_{\tau_i}^{\tau_f}\dd\tau \sqrt{-\dot{x}^2}\,,
\end{equation}
where $x^{\mu}=x^{\mu}(\tau)$ are $D$ real functions that specify the embedding in the target Minkowski space and $\tau$ is the parametrization of the curve. This action is reparametrization invariant and it can be easily generalized to any curved target space by simply replacing the Minkowski metric with a general Riemannian metric. Reparametrization invariance (local diffeomorphisms) implies that the canonical Hamiltonian vanishes and therefore the time evolution of the system is driven solely by constraints. The relativistic particle is a simple example of such singular constrained Hamiltonian system, since by definition of the canonical momentum conjugate to $x^{\mu}$, a single constrained is obtained, 
\begin{equation}
    \phi:=p^2 +m^2=0\,.
\end{equation}
This constraint is primary (it holds by definition of the canonical momentum and without the use of any equations of motion) and first class (trivially, since there is only one constraint.) Recall that in Dirac's classification first class constraints generate gauge symmetries (we refer to the textbook \cite{Henneaux:1994lbw} for a modern exposition.) For a suitable gauge choice, the Hamiltonian is proportional to the constraint, 
\begin{equation}
    H=\frac{p^2+m^2}{2m}\,,
\end{equation}
$\tau$ is the proper time and  $\dot{x}^2=-1$, the timelike trajectory of a massive particle. The Euler-Lagrange equations are simply Newton's second law $m\ddot{x}^{\mu}=0$. 

Famously, there are two reasons to seek an alternative classical action functional for the relativistic particle. The first is that the square root is hard to work with and in particular it does not facilitate quantization, and the second is that massless particles cannot be described by the previous action. To account for these issues, we couple the particle to an auxiliary field $e=e(\tau)$ (sometimes called an Einbein, in reference to the fact that in higher dimensions this coupling corresponds to a gravitational coupling of the particle.) The alternative action functional is 
\begin{equation}
    S'=\frac 12\int\dd\tau \, (e^{-1}\,\dot{x}^2-e \, m^2)\,. 
\end{equation}
It is also diffeomorphism invariant provided that in addition to the transformation of $x^{\mu}$, the Einbein transforms as a scalar density of weight $1$. The Euler-Lagrange equations for $e$ and $x^{\mu}$ are, respectively, 
\begin{equation}
    \dot{x}^{2}+e^2m^2=0 \quad \text{and} \quad \frac{\dd}{\dd\tau}(e^{-1}\dot{x}^{\mu})=0\,.
\end{equation}
The first equation is algebraic for $e$, which therefore may be eliminated from the action by solving for it in this equation. The resulting action after integrating out the field $e$ is precisely the original action $S$, and the two actions are classically equivalent.  It is worth mentioning that the new action does not have primary constraints, and that $\phi=0$ is a secondary constraint (valid by using the equation of motion), although it is still first class. Note now that in the new action we can use local diffeomorphism invariance to choose a gauge, in particular setting for $m\ne 0$ 
\begin{equation}
    e=\frac 1m \quad \Rightarrow \quad \ddot{x}^{\mu}=0 \quad \text{and} \quad \dot{x}^{2}=-1\,,
\end{equation}
describing a free massive particle, whereas for $m=0$ we set 
\begin{equation}
    e=1\quad \Rightarrow\quad \ddot{x}^{\mu}=0 \quad \text{and} \quad \dot{x}^{2}=0\,,
\end{equation}
the correct description for free massless particles in Minkowski spacetime. The generalization of this statement to curved spacetimes holds by replacement of the Minkowski metric with a (pseudo)Riemannian metric $g_{\mu\nu}$. 

\subsection{The Kawaguchi particle}
\label{subsec:Kawaguchi particle}

For the purposes of this paper, we would like to establish an analogous equivalence for higher derivative actions. We are specifically interested in the propagation of a free particle in a KL$^2$ spacetime, as described above, introducing a mass parameter $m$ for context,
\begin{equation}
    S_{\text{KL$^2$}}=m\int \dd\tau \, \sqrt[3]{\sfrac 12 \omega_{\mu\nu}\dot{x}^{\mu}\ddot{x}^{\nu}+\sfrac 16 \gamma_{\mu\nu\rho}\dot{x}^{\mu}\dot{x}^{\nu}\dot{x}^{\rho}}\,.
\end{equation}
As already stated, the action $S_{\text{KL$^2$}}$ is reparametrization invariant. As in the previous case, the original form of the action is only valid for nonzero mass and the cubic root is even more uncomfortable to handle. We can use the same trick of introducing the Einbein $e(\tau)$ to bring it into a more convenient form. The alternative action is 
\begin{equation}\label{eqA}
    S_{\text{KL$^2$}}'=\frac 13 \int\dd\tau \left(e^{-1}\,\sfrac 12 \omega_{\mu\nu}\dot{x}^{\mu}\mathring{a}^{\nu}+2\sqrt{e} \,m^{3/2}\right)\,,
\end{equation}
where the power of the mass in the second term in the parentheses guarantees correct dimensionality. The equation of motion for the auxiliary $e$ is algebraic, namely 
\begin{equation}
\sfrac 12 \omega_{\mu\nu}\dot{x}^{\mu}\mathring{a}^{\nu}-(em)^{3/2}=0\,,
\end{equation}
and eliminating the auxiliary field from the action, $S'_{\text{KL$^2$}}$ turns out to be completely equivalent to the original action $S_{\text{KL$^2$}}$ at the classical level. 
The action in Eq.~\eqref{eqA} exhibits local diffeomorphism invariance, provided that the Einbein transforms as a scalar density of weight 2. Indeed under the transformations
\begin{equation}
    \delta x^\mu=-\xi\dot{x}^\mu \qquad \text{and}\qquad  \delta e=-(\xi\dot{e}+2\,\dot{\xi}e)~,
\end{equation}
 the Lagrangian density transforms as a total derivative, namely $\delta L'=\frac{\partial}{\partial \tau}(-\xi L')$.
Gauge-fixing $e=1$ for $m=0$ results in the action for the massless Kawaguchi particle:
\begin{equation}
  S'_{\text{KL$^2$},0}=\frac 13\int \dd\tau \,\sfrac 12 \omega_{\mu\nu}\dot{x}^{\mu}\mathring{a}^{\nu}=\frac 13 \int \dd\tau\,\left(\sfrac 12 \omega_{\mu\nu}\dot{x}^{\mu}\ddot{x}^{\nu}+\sfrac 16\gamma_{\mu\nu\rho}\dot{x}^{\mu}\dot{x}^{\nu}\dot{x}^{\rho}\right)\,.  
\end{equation}
Up to an irrelevant overall numerical factor, this is precisely the action we were looking for, since it agrees with the $\epsilon^3$ sector of the effective action in Eq.~\eqref{eq:action:3rdorder}. 

We can now write down the Euler-Lagrange equations for this action, using the general formula \eqref{eq:EL higher general formula}, as before. 
A straightforward computation reveals that the Euler-Lagrange equations for $S_{\text{KL$^2$},0}'$ are
\begin{equation}
  \omega_{\mu\nu}\dddot{x}^{\nu}+(\gamma_{\mu\kappa\lambda}-\sfrac 12 H_{\mu\kappa\lambda}-2\partial_{(\kappa}\omega_{\lambda)\mu})\dot{x}^{\kappa}\ddot{x}^{\lambda}-\sfrac 16(\partial_{\mu}\gamma_{\kappa\lambda\rho}-3 \partial_{\kappa}\gamma_{\mu\lambda\rho}+3\partial_{\nu}\partial_{\rho}\omega_{\kappa\mu})\dot{x}^{\kappa}\dot{x}^{\lambda}\dot{x}^{\rho}=0\,,
\end{equation}
i.e. the $\epsilon^3$ part of the effective equations of motion in Eq.~\eqref{eq:EL:3}.
This is essentially the geodesic equation for KL$^2$ geometry. We will discuss its geometrical interpretation in terms of a covariant jerk vector in the Section \ref{sec:jerk}. 

We note that to combine the statements of Sections \ref{subsec:relativistic particle} and \ref{subsec:Kawaguchi particle} we would need two different auxiliary fields $e$ and $e'$, one to eliminate the square root and one to eliminate the cubic root. This is due to the fact that the auxiliary field in each case transforms as a scalar density of different weight (1 and 2, respectively) under coordinate transformations. 

\section{Jerk vector and manifest covariance}
\label{sec:jerk}

Our purpose now is to give a geometrical perspective on the third order effective equation of motion \eqref{eq:EL:3}. Apart from the terms proportional to $\epsilon^3$, everything is already in a manifestly covariant form. Inspecting the third order terms, we observe that the first term contains the third time derivative of position, called jerk. 
The jerk, and for that matter any other higher time derivative, is similar to the acceleration in that it requires additional structure to be properly defined in curved space, i.e.~such that it transforms as a vector. 

Recall that the acceleration vector in a curved space is defined through the parallel transport of the velocity, which is equivalent to saying that the acceleration is the covariant derivative of the velocity. Similarly, the jerk is defined as the covariant  derivative of the acceleration.
To keep the discussion general and given that the third order terms 
do not contain a metric, we recall that a metric is not necessary for parallel transport. The equation defining straight lines in a general affine space, the autoparallel equation, is
\begin{equation}
\label{eq:a}
    a^{\mu}:= \ddot{x}^{\mu}+\Gamma^{\mu}_{\kappa\lambda}\dot{x}^{\kappa}\dot{x}^{\lambda}=0\,,
\end{equation}
where $\Gamma^{\mu}_{\k\l}$ are the coefficients of an affine connection $\nabla$.  With respect to the velocity vector $v=\dot{x}(t)$ associated to a curve $x: \R\to \mc{M}$, we write $a=\nabla_{v}v$. The geodesic equation, on the other hand, is obtained when we consider a Riemannian metric $g_{\mu\nu}$, and, as we have already discussed, it is 
\begin{equation}
    \mathring{a}^{\mu}[g]:= \ddot{x}^{\mu}+\mathring{\Gamma}^{\mu}_{\kappa\lambda}[g]\dot{x}^{\kappa}\dot{x}^{\lambda}=0\,.
\end{equation}
When we consider spaces admitting an affine connection that is metric compatible and torsion-free, the autoparallel equation and the geodesic equation are identical. In general, however, an affine connection may be expressed as 
\begin{equation}
    \Gamma^{\mu}_{\nu\rho}=\mathring{\Gamma}^{\mu}_{\nu\rho}+K^{\mu}_{\nu\rho}+L^{\mu}_{\nu\rho}\,,
\end{equation}
where the tensors $K$ and $L$ are the contorsion and distorsion tensors defined as 
\begin{align}
K^{\mu}_{\nu\rho} & = \frac 12 g^{\mu\sigma}(T_{\nu\sigma\rho}+T_{\rho\sigma\nu}-T_{\nu\rho\sigma})\,,\\[4pt]
L^{\mu}_{\nu\rho} &= \frac 12 \left(Q^{\mu}{}_{\nu\rho}+Q^{\mu}{}_{\rho\nu}-Q_{\nu\rho}{}^{\mu}\right)\,, 
\end{align}
where $T$ is the torsion tensor with components $T^{\mu}_{\nu\rho}=\Gamma^{\mu}_{\nu\rho}-\Gamma^{\mu}_{\rho\nu}$ and $Q$ is the nonmetricity tensor with components $Q_{\mu\nu\rho}:=-\nabla_{\mu}g_{\nu\rho}$.
In spaces with torsion and/or nonmetricity, the autoparallel and geodesic equations are different, and so are the definitions of the acceleration vector.  

When defining the jerk through the parallel transport of the acceleration vector, we may want to use a different connection than the one used to define the acceleration. This might be the case if we have additional geometric structure in the problem, for example a symplectic or an almost symplectic structure, as is the case in the problem we are dealing with in this paper. In the present context, focusing only on the cubic terms we have the symmetric connection $\mathring\nabla$ defined through Eq.~\eqref{eq:def of gamma}, which gave rise to the vector $\mathring{a}^{\mu}$, which is different from the vector $a^{\star\mu}$ in the effective equation of motion \eqref{eq:EL:3} defined in turn through the Levi-Civita connection for the renormalized mass tensor. Suppose then that we have another linear connection $\nabla$, possibly with nonvanishing torsion. We define the jerk vector to be 
\begin{equation}\label{eq:jerk}
    j:=\nabla_v\mathring{a}= \nabla_v\mathring\nabla_{v}\,v\,.
\end{equation}
If $\mathring\Gamma$ and $\Gamma$ are the coefficients of $\mathring\nabla$ and $\nabla$ respectively, the component expression of the jerk vector is found to be 
\begin{equation}
\label{eq:jerk components}
    {j}^{\mu}=\dddot{x}^{\mu} +\ddot{x}^{\kappa}\dot{x}^{\lambda}(2\,\mathring\Gamma_{\kappa\lambda}^{\mu}+{\Gamma}_{\lambda\kappa}^{\mu})+\dot{x}^{\rho}\dot{x}^{\kappa}\dot{x}^{\lambda}(\partial_{\rho}\mathring\Gamma_{\kappa\lambda}^{\mu}+\mathring\Gamma_{\kappa\lambda}^{\nu}{\Gamma}_{\rho\nu}^{\mu})\,.
\end{equation}
This is the general form of the jerk vector on a Kawaguchi manifold of order 2. Obviously, the formula also holds in the special case $\nabla=\mathring\nabla$.

We can now use the definition of the jerk vector to express the effective equation of motion \eqref{eq:EL:3} in manifestly covariant form. First, solving Eq.~\eqref{eq:def of gamma} for the symmetric connection, 
\begin{equation}
\mathring\Gamma^{\mu}_{\nu\rho}=\sfrac 13 \omega^{\mu\sigma}\gamma_{\sigma\nu\rho}-\sfrac 23 \omega^{\mu\sigma}\partial_{(\nu}\omega_{\rho)\sigma}\,.
\end{equation}
Furthermore, inspecting the form of the function $\zeta_{\mu\nu\rho}$ in Eq.~\eqref{eq:zeta} we are prompted to make the following choice for the connection $\nabla$: 
\begin{equation}
    \Gamma^{\mu}_{\nu\rho}=\mathring\Gamma^{\mu}_{\nu\rho}-\sfrac 12 \,\omega^{\mu\sigma}H_{\sigma\nu\rho}\,.
\end{equation}
We note that this is not a symmetric connection. Its torsion is found to be 
\begin{equation}
    T^{\mu}_{\nu\rho}= -\, \omega^{\mu\sigma}H_{\sigma\nu\rho}\,.
\end{equation}
Moreover, its action on the almost symplectic form is once again proportional to $\dd\omega$, albeit with a different proportionality constant from Eq.~\eqref{eq:almost compatible connection}:{\footnote{We note that connections which are compatible with the almost symplectic structure, namely $\nabla\omega=0$ also exist. The torsion of a compatible connection must be equal to $-\sfrac 13\,\omega^{\mu\sigma}H_{\sigma\kappa\lambda}$.}} 
\begin{equation}
    \nabla_{\mu}\omega_{\nu\rho}=-\,\sfrac 23 H_{\mu\nu\rho}\,.
\end{equation}
Inserting these connection coefficients in the definition of the jerk vector, we find that the effective equation of motion \eqref{eq:EL:3} takes the manifestly covariant form
\begin{equation}
\label{eq:effective eom manifest covariance}
\epsilon^3 \omega_{\mu\nu}j^{\nu}-\sfrac 12\epsilon^3\omega_{\sigma\rho}\mathring{R}^\rho{}_{\kappa\lambda\mu}\dot{x}^\kappa\dot{x}^\lambda\dot{x}^\rho+M_{\mu\nu}^{\star}a^{\star\nu}-\epsilon B_{\mu\nu}\dot{x}^{\nu}+\partial_{\mu}E_{n}=0\,,
\end{equation}
where $\mathring{R}:= R_{\mathring\nabla}$ is the curvature of the connection $\mathring\nabla$. 
In this manifestly covariant form of the effective equation of motion only two third order terms remain. The first contains the jerk vector and the other contains the Riemann curvature tensor.  Despite being odd in time derivatives, both terms are invariant under time reversal, since they contain the antisymmetric tensor $\omega_{\m\n}$, which changes sign under time reversal, like the Berry curvature $B_{\m\n}$.  

We emphasize again that the appearance of only two terms in the cubic sector of the equations of motion is due to the particular choice of connection $\nabla$. Given that the difference of two connections is an endomorphism-valued 1-form, say $\Phi$, suppose we choose a different connection $\nabla'=\nabla+\Phi$. 
The new covariant jerk vector takes the form 
\begin{equation}
    j'^{\mu}=j^{\mu}+\Phi^{\mu}_{\nu\rho}(x)\dot{x}^{\nu}\mathring{a}^{\rho}\,.
\end{equation}
This means that a more general form of the cubic sector of the equation of motion is 
\begin{equation}
\label{eq:cubic:general}
    \omega_{\mu\nu}j'^{\nu}+\omega_{\mu\nu}\Phi^{\nu}_{\rho\sigma}\dot{x}^{\rho}\mathring{a}^{\sigma}-\sfrac 12 \omega_{\sigma\rho}\mathring{R}^{\rho}{}_{\kappa\lambda\mu}\dot{x}^{\kappa}\dot{x}^{\lambda}\dot{x}^{\rho}=0\,.
\end{equation}
Note that this new connection satisfies 
\begin{equation}
    \nabla'_{\mu}\omega_{\nu\rho}=-\sfrac 23 H_{\mu\nu\rho}+2\Phi^{\sigma}_{\mu[\nu}\omega_{\rho]\sigma}\,. 
\end{equation}
Observe that there exists a connection satisfying the compatibility condition $\nabla'\omega=0$ upon choosing $\Phi^{\mu}_{\nu\rho}=\sfrac 13 \omega^{\mu\kappa}H_{\kappa\nu\rho}$. For this choice, there is a term $\sfrac 13 H_{\mu\rho\sigma}\dot{x}^{\rho}\mathring{a}^{\sigma}$ in the equation of motion. 
Equation \eqref{eq:cubic:general} is not yet the most general form of the cubic terms. A completely general expression is obtained once a difference form is considered also for the second connection $\mathring{\nabla}$ used in defining the jerk. Since this does not add anything substantial to our analysis, we do not present these general expressions here. 

As a final note, when a connection with vanishing curvature exists globally, we call the manifold a flat  
KL$^2$ manifold. This is not the case in general, since the existence of a flat connection imposes topological restrictions on the manifold. However, when this condition is met, the effective equation of the Kawaguchi action turns out to be simply the vanishing of the jerk vector, $j^{\mu}=0$.

\section{Conclusions} 
\label{sec:conclusions}

When the dynamics in a many-variable system takes place on two distinct time scales, the complexity of the coupled equations of motion for the fast and slow variables is often reducible to the complexity of self-contained equations of motion for the slow variables.  This simplification is possible because the fast variables become adiabatically locked to the slow variables and can therefore be eliminated in favor of the slow variables.  The resulting effective equation of motion for the slow variables contains a term which can be interpreted as a force if the slow variables are classical.

Quantum-classical molecular dynamics provides an illustrative example. Eliminating the electronic degrees of freedom yields an effective equation of motion for the classical nuclei with an effective force that can be expanded in powers of the adiabatic parameter.  The zeroth order force is the gradient of the adiabatic potential energy surface.  The first order force is the effective Berry-Lorentz force of geometric magnetism.

We used adiabatic perturbation theory to calculate higher order forces.  The second order force produces a nonadiabatic mass tensor, which renormalizes the bare nuclear mass tensor, causing the nuclei to move as if they lived in a curved space with a different Riemannian metric.  The second order equation is a forced geodesic equation.
It is worth emphasizing that the dimension of the classical parameter space (nuclear configuration space) is not assumed to be equal to the dimension of the (electronic) Hilbert space and may be lower than the latter.  In the present work, we focus on even-dimensional parameter spaces, although in general they can be even- or odd-dimensional.

At third order, the equation of motion is no longer Newtonian because the adiabatic reaction force depends on the jerk, the third order time derivative of position.  Nevertheless, it is still fundamentally geometrical and our derivation shows that it is the Euler-Lagrange equation of an effective action functional depending on second order time derivatives. We found that the third order effective dynamics is ruled by Kawaguchian geometry instead of Riemannian geometry.  Just as the second order terms were equivalent to the geodesic equation in a Riemannian space, the third order terms define a generalized geodesic in a type of Kawaguchi space where the differential arc length depends on the acceleration, not just on the velocity. 

This Kawaguchi space, for which we used the terminology KL$^2$ space, is characterized by an almost symplectic structure $\omega_{\mu\nu}$ (a nondegenerate 2-form) and a symmetric quantity $\gamma_{\mu\nu\rho}$ that obeys an affine transformation law under a change of coordinates. Together these structures give rise to a connection, which is not unique but assisted us in expressing the equation of motion in a manifestly covariant form. This became possible once we identified a covariant jerk vector, since the third order time derivative of position has no coordinate invariant meaning in a non-Euclidean space. 

In the particular example of quantum-classical molecular dynamics, we gave explicit formulas for $\omega_{\m\n}$ and $\gamma_{\m\n\r}$ in terms of the adiabatic electronic state and its derivatives with respect to the nuclear parameters.  The almost symplectic structure is the imaginary part of a new Hermitian metric, an analog of the quantum geometric tensor containing the square of the resolvent operator, which arises at third order in adiabatic perturbation theory.  The symmetric quantity $\gamma_{\mu\nu\rho}$ has no known quantum geometric analog. 

The second and third order adiabatic reaction forces can be implemented in molecular dynamics simulations to capture new phenomena. Despite being third order in $\e$, the third order force can have a qualitative effect on the long-time dynamics.  This could occur when the nuclear motion has an oscillatory component superimposed on a slow ``drifting'' component.  
If the time average of the effective mass tensor $M^{\star}_{\m\n}$ over a period of oscillation is a flat metric while the time average of the third order force is nonzero, the third order kicks can build up to a qualitative effect on long time scales.
This is similar to how the rotation of the Foucault pendulum becomes visible at long times because the Coriolis force, despite being weak, has a nonvanishing time average, while the much larger gravitational restoring force averages out.

Variational problems where the energy functional depends on the covariant acceleration or higher covariant derivatives thereof are also important in the context of obstacle avoidance where configuration spaces are Riemannian manifolds and in control theory and interpolation problems, see e.g.~Refs.~\cite{MC,MC2} and references therein for a variety of applications in physics and engineering. However, in those works the equations of motion contain even order derivatives, precisely because of the Riemannian metric, whereas equations of motion with odd order derivatives fall through the cracks. The effective geometry that we uncovered in this paper, which involves antisymmetric tensors like the almost symplectic structure $\omega$, remedies this and includes equations with odd order derivatives, for example through the jerk vector that we encountered at third order in adiabatic perturbation theory. 

The jerk also plays a fundamental role in higher-derivative Lagrangians used to describe the motion and behavior of biotechnological (e.g.~robotic arms) and even biological (e.g.~human motion, neural control) systems \cite{balasubramanian_analysis_2015,Todorov,Boulanger_2024,Nicolas2}. Once again though, these Lagrangians are quadratic in the jerk, unlike the one we derived in this paper. It would be interesting to explore potential applications of our approach in these directions, beyond the realm of molecular dynamics.  

\paragraph{Acknowledgements.} The authors thank Jan Rosseel for discussions and participation in the early stages of this work. We thank Dimitri Giataganas for discussions and Dmitry Roytenberg for pointing out Ref.~\cite{Chern}. During the completion of this work, the authors received funding from the Croatian Science Foundation projects ``Higher Structures and Symmetries in Gauge and Gravity Theories'' (IP-2024-05-7921) and ``Mining the Quantum: Frustration, Disorder, and Devices'' (IP-2025-02-1667) and from the European Union -- NextGenerationEU.

\bibliography{refs}
\bibliographystyle{unsrt}

\end{document}